\documentclass[aps,prx,longbibliography,showpacs,twocolumn]{revtex4-1}
\usepackage{CDArticle}
\usepackage{times}

\usepackage{amssymb}
\usepackage{epsfig}
\usepackage{graphicx}
\usepackage{amsmath}
\usepackage{array,color}
\usepackage{natbib}

\newcommand{\ve}[1]{\mathbf{ #1}}							

\newcommand{\TC}{T_\text c}				

\newcommand{\tg}[1]{\textbf{#1}}     

\newcommand{\defi}[2]{\stackrel{{#1}}{#2}}			

\newcommand{\BEI}{\begin{IEEEeqnarray}{rCl}}
\newcommand{\EEI}{\end{IEEEeqnarray}}
\newcommand{\BE}{\begin{equation}}
\newcommand{\BEA}{\begin{eqnarray}}
\newcommand{\EE}{\end{equation}}
\newcommand{\EEA}{\end{eqnarray}}
\newcommand{\NN}{\nonumber}

\newcommand{\UU}{$U/t$ }
\newcommand{\BEDT}{$\kappa$-(ET)$_{2}$X }
\newcommand{\KCN}{$\kappa$-(ET)$_{2}$Cu$_{2}$(CN)$_{3}$ }
\newcommand{\KCl}{$\kappa$-(ET)$_{2}$Cu[N(CN)$_{2}$]Cl }
\newcommand{\DMITS}{Pd(dmit)$_{2}$}

\newcommand{\OP}{$d_{SC}$ }
\newcommand{\TCM}{$T_{c}^{\, m}$ }
\newcommand{\UCHI}{$\chi_{z}$ }

\begin{document}
\title{Superconducting dome in doped quasi-2d organic Mott insulators: a paradigm for strongly-correlated superconductivity}
\author{Charles-David H\'{e}bert$^1$}
\author{Patrick S\'{e}mon$^1$}
\author{A. -M. S. Tremblay$^{1,2}$}
\affiliation{$^1$D\'{e}partement de physique and Regroupement qu\'{e}b\'{e}cois sur les mat\'{e}riaux de pointe, Universit\'{e} de Sherbrooke, Sherbrooke, Qu\'{e}bec, Canada, J1K 2R1\\
	$^2$Canadian Institute for Advanced Research, Toronto, Ontario, M5G 1Z8, Canada
	}
\date{\today }
\email{charles-david.hebert@usherbrooke.ca \\ andre-marie.tremblay@usherbrooke.ca}

\begin{abstract}
Layered organic superconductors of the BEDT family are model systems for understanding the interplay of the Mott transition with superconductivity, magnetic order and frustration, ingredients that are essential to understand superconductivity also in the cuprate high-temperature superconductors. Recent experimental studies on a hole-doped version of the organic compounds reveals an enhancement of superconductivity and a rapid crossover between two different conducting phases above the superconducting dome. One of these phases is a Fermi liquid, the other not. Using plaquette cellular dynamical mean field theory with state of the art continuous-time quantum Monte Carlo calculations, we study this problem with the two-dimensional Hubbard model on the anisotropic triangular lattice. 
Phase diagrams as a function of temperature $T$ and interaction strength \UU are obtained for anisotropy parameters $t'=0.4t$, $t'=0.8t$ and for various fillings.
As in the case of the cuprates, we find, at finite doping, a first-order transition between two normal-state phases. One of theses phases has a pseudogap while the other does not. At temperatures above the critical point of the first-order transition, there is a Widom line where crossovers occur.  The maximum (optimal) superconducting critical temperature \TCM at finite doping is enhanced by about $25\%$ compared with its maximum at half-filling and the range of \UU where superconductivity appears is greatly extended. These results are in broad agreement with experiment. Also, increasing frustration (larger $t'/t$) significantly reduces magnetic ordering, as expected. This suggests that for compounds with intermediate to high frustration, very light-doping should reveal the influence of the first-order transition and associated crossovers. These crossovers could possibly be  even visible in the superconducting phase through subtle signatures. We also predict that destroying the superconducting phase by a magnetic field should reveal the first-order transition between metal and pseudogap. Finally, we predict that electron-doping should also lead to an increased range of \UU for superconductivity but with a reduced maximum $\TC$. This work also clearly shows that the superconducting dome in organic superconductors is tied to the Mott transition and its continuation as a transition separating pseudogap phase from correlated metal in doped compounds, as in the cuprates. Contrary to heavy fermions for example, the maximum $\TC$ is definitely not attached to an antiferromagnetic quantum critical point. That can also be verified experimentally.  
\end{abstract}

\maketitle




\section{Introduction}


\paragraph*{} 
 In organic charge transfer salts, such as $\kappa \text{-(BEDT-TTF)}_{2}X$ (\BEDT) or  $\text{Et}_{n}\text{Me}_{4-n}\text{Pn[Pd(dmit)}_{2}]_{2}$   (\DMITS), a first-order phase transition between a superconductor and a Mott insulator is induced by pressure \cite{Lefebvre_Phase_Diagram_2000, Shimizu_Pressure_Induced_2010, Shimizu_Dmit, Yamaura_Dmit,Limelette_Mott_2003}. These materials also present a wide range of intriguing phenomena such as unconventional superconductivity, magnetic ordering \cite{Limelette_Mott_2003}, pseudogap \cite{Merino_Pseudogap_2014} valence-bond solid phases \cite{Shimizu_Dmit} and some of them are even spin-liquid candidates. ~\cite{Shimizu_SpinLiquid_2003, Yamashita_Highly_DMIT_2010}. Excellent reviews are available.~\cite{powell_strong_2006,Powell_Quantum_2011}

The presence of the Mott transition and of spin-liquid states in the phase diagram suggests that strong electronic correlations and electronic frustration are key to the physics of the organics. The one-band Hubbard model on an anisotropic triangular lattice near half-filling is the simplest model that captures this physics~\cite{KinoFukuyama:1996,powell_strong_2006,Powell_Quantum_2011}, although consensus has not yet been completely reached~\cite{Clay:2008,Dayal_Absence_2012} on this point. 
 
Unraveling the physics of these layered materials should also be helpful to shed light on cuprate high-temperature superconductors. Indeed, these two classes of materials give a complementary perspective on the crucial role of the Mott transition. In the organics, the Mott transition is bandwidth-controlled whereas it is doping-controlled in the cuprates.~\cite{Imada_Rev_Mod_Phy,TremblayJulichPavarini:2013}  The analogy between these two classes of materials has been reinforced through recent experimental studies~\cite{Oike_Mottness_2014} of doped organics~\cite{Lyubovskaya-1:1987,Lyubovskaya-2:1987} that show a superconducting dome~\cite{Budko:1987} as a function of pressure as well as a rapid change from non-Fermi liquid (pseudogap) to Fermi-liquid like metal at a critical pressure in the normal state.~\cite{taniguchi_anomalous_2007,Oike_Mottness_2014} These analogies motivate our work. In short, our calculations explain these different features just as calculations performed with the same methods explain many of the key features of the cuprate phase diagram.~\cite{Taillefer:2015} In particular a first-order transition at finite doping and its associated Widom line play a crucial role as in the cuprates.~\cite{Sordi_Widom,Sordi_c_axis_2013}  We define superconductivity as strongly-correlated when it arises in the presence of interactions larger than, or of the order of, those necessary to lead to a Mott transition at half-filling. 

Some of the striking experimental results that we address and understand theoretically in this paper come from recent work on $\kappa$-(ET)$_{4}$Hg$_{2.89}$Br$_{8}$ by Oike {\it et al.}~\cite{Oike_Mottness_2014}. That compound is considered as a 10\% doped analog of \KCN. The main observation is that maximally enhanced superconductivity and a normal-state crossover to a non-Fermi liquid phase appear concomitantly around a pressure where mobile carriers decrease rapidly. Also, the range of pressures spanned by the superconducting dome in $\kappa$-(ET)$_{4}$Hg$_{2.89}$Br$_{8}$ is about six times the range where it appears in the half-filled analog \KCN. 

\paragraph*{} 
We work with the Hubbard model on the anisotropic triangular lattice as a function of temperature, interaction strength \UU and filling $n$, for different values of frustration characterized by the ratio of near-neighbor hopping $t'/t$ in different directions. Values for $t'/t$ are inspired from Kandpal  \textit{et al.} \cite{Kandpal_Revision_Parameters_2009} and from Nakamura  \textit{et al.}~\cite{Nakamura_Revision_Parameters_2009} who found, using \textit{ab initio} density functional theory, that \KCl could be modeled by $t'=0.4t$ and \KCN by $t'=0.8t$. Comparing the results for two values of $t'/t$ helps understand the effect of magnetic frustration on the phase diagram.

We use a cluster generalization of dynamical mean-field theory.~\cite{Georges:1996,Maier_Cluster_Size_SC,KotliarRMP:2006,LTP:2006} This approach has already led to numerous results that  can be confronted with experiments, thus  permitting to address important issues in cuprates, \DMITS and \BEDT such as the pseudogap, superconductivity, Mott transition, magnetic ordering, thermodynamic properties, and unusual criticality in organic compounds.~\cite{TremblayJulichPavarini:2013} The assumption inherent in our approach is that the main physics of the organics originates from strong correlations that occur at short distances due to on-site repulsion $U$ and near-neighbor superexchange $J$ that are present in the Hubbard model. This is the assumption behind the state of the art method that we use for a $2\times 2$ cluster embedded in a self-consistent dynamical mean-field. Larger cluster calculations would be necessary if this assumption was proven incorrect.  The agreement between our results at low temperature and those obtained recently through variational Monte Carlo methods helps establish the validity of the approach.~\cite{Watanabe_SC_2014} Agreement with several experimental facts strongly suggests that our approach is relevant for experiment. We make predictions for experiment that can falsify the theory. In the absence of an exact solution to the Hubbard model, we are making a minimum number of assumptions and suggesting experiments that can falsify them by disagreeing with the consequences. 

\begin{figure} \label{fig:GeneralizedPhaseDiagram} 
	\includegraphics[width=\linewidth]{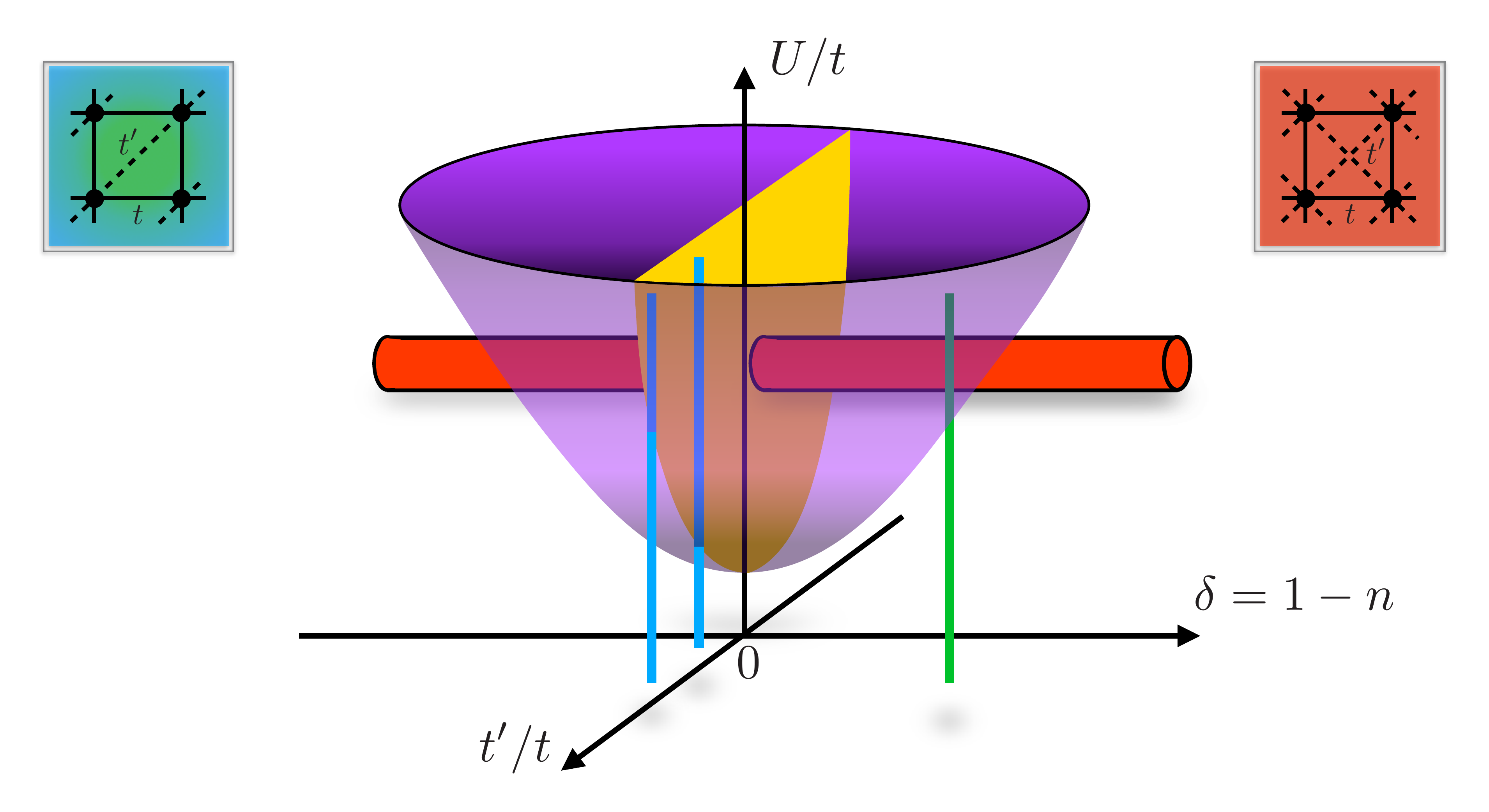}	
	\caption{Schematic generalized normal state phase diagram for the cuprates and the layered organics, in the limit of low temperature neglecting broken-symmetry phases. The yellow surface represents the Mott insulating phase at zero hole-doping $\delta=0$, or half-filling $n=1$. Leaving the Mott insulating phase by increasing doping yields a second-order transition to a pseudogap phase up to the magenta surface where 
	a first-order transition to a correlated metal occurs (at least close to half-filling). That first-order transition extends from the first-order Mott metal-insulator transition that occurs at the boundary of the yellow region. Cuprate superconductors are found in the red region at negative $t^\prime/t$. The half-filled layered organics are found for different $t^\prime/t$ along the blue regions in the zero hole-doping plane. They are strongly influenced by pressure, whose effect is represented vertically, although the ratio $t^\prime/t$ will generally also be influenced by pressure.  A {\it doped} layered organic is found in the green region. It allows one to expand the analogy between organics and cuprates. The organic and cuprate lattices are different, as illustrated respectively by the left and right insets. Nevertheless, the physics of interactions, frustration and doping is present in both types of compounds. Superconductivity and antiferromagnetism are broken symmetry phases that are strongly influenced by the underlying normal state illustrated by this figure. In particular, the superconducting phase has a maximum $T_c$ in the vicinity of the boundary of the magenta region.}
\end{figure}

The link between the normal state of high-temperature cuprate superconductors and that of the organics is illustrated schematically~\footnote{We thank A. Reymbaut for drawing this figure.} in Fig.~\ref{fig:GeneralizedPhaseDiagram}. Disregarding temperature, the relevant variables are interaction strength \UU, doping $\delta$, and frustration as measured by the ratio of second to first nearest-neighbor hopping $t^\prime/t$. Cuprate superconductors, in the red region, are easily doped but are little influenced by pressure and the range of $t^\prime/t$ varies little between different compounds. Layered organic compounds on the other hand are half-filled, with a broad range of possible values of frustration $t^\prime/t$ from compound to compound, and their bandwidth to interaction ratio is strongly influenced by pressure. A pseudogap appears through a second-order transition when doping is increased from the yellow region containing the Mott insulating phase at half-filling. The pseudogap phase ends in a first-order manner on the magenta surface in Fig.~\ref{fig:GeneralizedPhaseDiagram}. The latter first order transition extends from the first-order Mott insulator-metal transition occurring at half-filling at the boundary of the yellow region.

The range of parameters that could be relevant for the doped organic compound $\kappa$-(ET)$_{4}$Hg$_{2.89}$Br$_{8}$  is indicated by the green region in Fig.~\ref{fig:GeneralizedPhaseDiagram}. This compound then offers the interesting possibility to investigate the pseudogap to correlated metal transition in the normal state by cutting the first-order magenta surface along a direction different from that of the cuprates. The effect of this transition on the superconducting state is also a key question that we address here. 

\paragraph*{} 
The Hubbard model and the Cellular Dynamical Mean-Field Theory on a plaquette with Continuous-Time Quantum-Monte-Carlo  impurity solver are presented in Sec.~\ref{sec:ModelandMethod}. This work would not have been possible without recent improvements of this solver related to sign problem minimization,~\cite{Semon_Importance_2012} ergodicity~\cite{Semon_Ergodicity_2014}, and speedup.~\cite{Semon_Lazy_2014} We begin in Sec.~\ref{Sec:cuprates} with a short summary of some previous results for the cuprates. We then present in Sec.~\ref{Sec:normal} results for the normal state, showing the Widom line that emerges from the first-order transition~\cite{Sordi_Finite_Doping_2010,Sordi_Metal_Transitions_2011} between a pseudogap phase and a metal.~\cite{Sordi_Widom} This plays a crucial role for the cuprates.~\cite{Taillefer:2015} The results in Sec.~\ref{sec:results} are for two different lattice anisotropies, $t'/t$, or equivalently, frustration. We investigate the N\'{e}el antiferromagnetism (AFM) and d-wave superconductivity (SC) on the same footing but the relative stability of the phases is not studied. For $t'/t=0.4$, half-filling, 1\%, 10\% hole-doping and 10\% electron-doping are investigated.  We find for a $1\%$ hole-doping that the maximum (optimal) superconducting  critical temperature (\TCM) as a function of pressure or (interaction strength) is enhanced by approximately $25\%$ and the range of superconductivity is multiplied by a factor of six on the pressure axis ($t/U$). The range of pressure where superconductivity exists for $10\%$ doping is similar. We also obtain the \TCM line in the $T-U-n$ phase diagram. The case $t'/t=0.8$ is considered only for the 10\% hole-doped case and at half-filling due to a worse sign problem. Discussions in Sec.~\ref{sec:discussion} include the role of long wave-length fluctuations on broken-symmetry phases, the role of antiferromagnetic quantum critical points and of the Mott transition on the superconducting dome, contact with experiment, predictions, limitations of the approach, and perspectives.  The most important conclusions are summarized in Sec.~\ref{sec:conclusion}.


\begin{figure}[!htbp]
\includegraphics[width=0.5\linewidth]{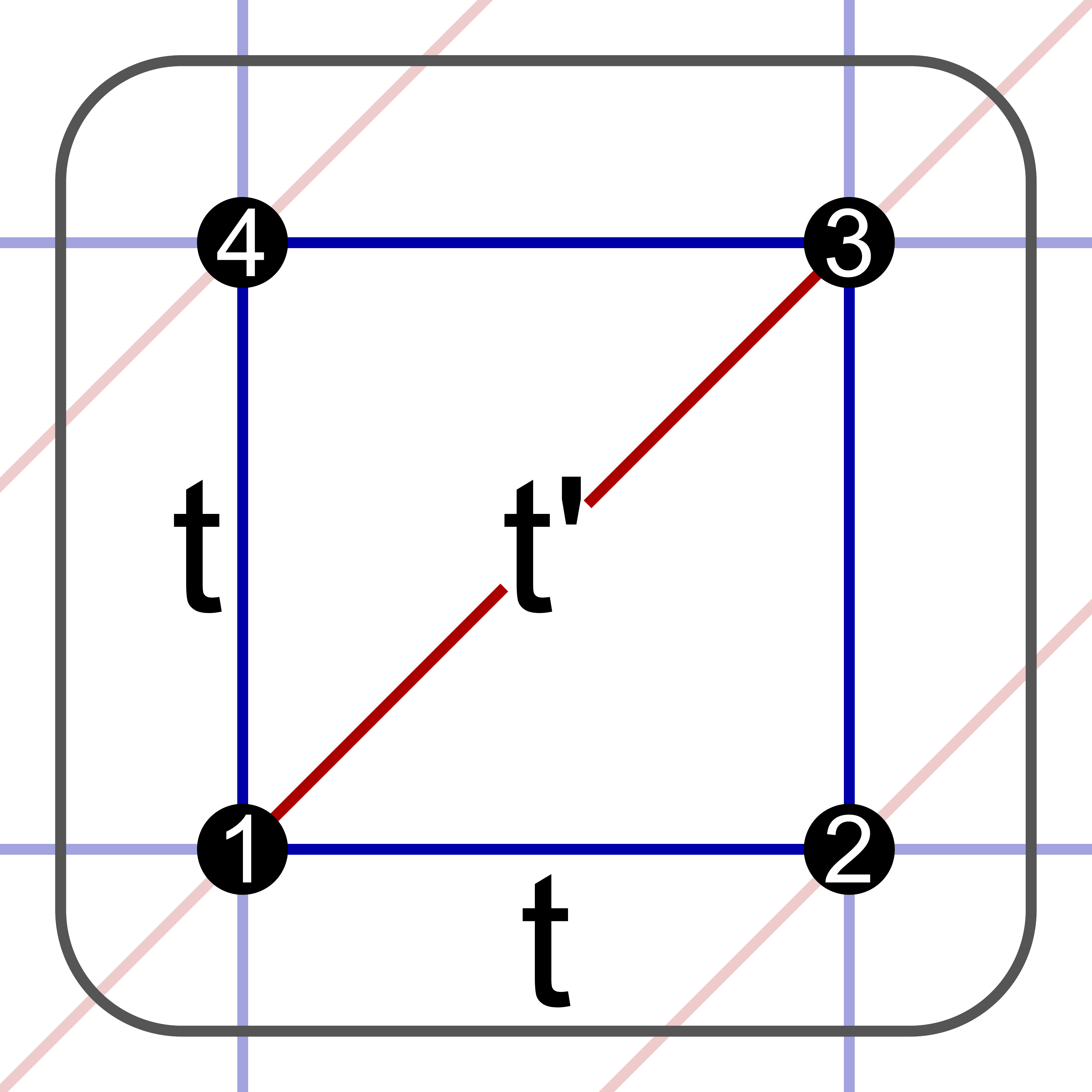}
\caption{Periodic partitioning of the anisotropic triangular lattice into $2 \times 2$ frustrated square clusters for this work using CDMFT.
}
\label{fig:AnisotropicSquareLattice}
\end{figure}

\section{Model and method}\label{sec:ModelandMethod}
We consider the single band Hubbard Hamiltonian on the anisotropic triangular lattice in two dimensions 

\BEI
\label{eq:Hubbard}
H & = & \sum_{i,j,\sigma}t_{ij} \, c^\dagger_{i\sigma}c_{j\sigma}+ U \sum_i n_{i\uparrow} n_{i\downarrow}-\mu\sum_{i,\sigma}n_{i,\sigma} 
\EEI
where $t_{ij}$ is the hopping amplitude between neighboring sites, $c_{i\sigma}$ and $c^{\dagger}_{i\sigma}$ respectively
destroy and create an electron of spin $\sigma$ at site $i$,
$n_{i\sigma}$ is the density of electrons of spin $\sigma$ at site i, $\mu$ is the chemical potential
and $U$ is the on-site Coulomb repulsion.  This model was proposed several decades ago for the organics.~\cite{KinoFukuyama:1996} Its validity has been revisited recently for specific compounds.~\cite{Nakamura:2009} In particular, it has been argued that near-neighbor repulsion $V$ was important. Since previous studies for the cuprates have shown that $V$ does not influence the phase diagram in a dramatic way in the strong-correlation limit,~\cite{Next_SC_Senchal} we neglect this term in this initial study. 

As illustrated in Fig.~\ref{fig:AnisotropicSquareLattice}, we take $t_{ij}=t$ for nearest-neighbor bonds, and $t_{ij}=t'$ for the diagonal bond. The isotropic triangle is recovered for $t'=t$. The ratio $t'/t$ is both an anisotropy parameter and a measure of magnetic frustration. The two expressions are used interchangeably. In the figures, inverse temperature $\beta$ is given in units of $1/t$. 

$T_c$ stands for the superconducting critical temperature. It was called $T_c^d$ in Ref.~\onlinecite{Sordi_Strong_2012} to emphasize that it is the dynamical mean-field transition temperature, that differs from the true superconducting transition temperature. We use \TCM for the maximum value that this quantity takes at a given doping as a function of pressure or interaction strength.

\subsection{CDMFT}

\paragraph*{}
In two dimensions, momentum-dependence of the self-energy is important. Cellular dynamical mean field theory~\cite{Kotliar_Cellular_2001} (CDMFT) for the Hubbard model takes into account short-range correlations in addition to interaction-induced dynamical correlations; single-site DMFT~\cite{Georges:1996} is not appropriate to study the momentum-dependence associated to d-wave superconductivity. The key approximation is to restrict the self-energy to a local cluster and neglect its spatial dependence beyond the cluster~\cite{Maier_Quantum_2005, KotliarRMP:2006,LTP:2006,Senechal_Introduction_2008}.

In practice, CDMFT embeds a cluster of finite size in a non-interacting electronic bath. The impurity problem (cluster and bath) is then solved and the bath is determined self-consistently by demanding that the lattice Green function projected on the cluster equal the Green function obtained from the impurity problem.

To be more specific, the lattice Green function in Matsubara frequencies is obtained from \BEI
\label{eq:GreenLattice}
\hat{G}_{latt}^{-1}(i\omega_n,\defi{\sim}{\ve{k}}) & = & \left( i\omega_n + \mu \right)\hat{\mathbb{I}} -\hat{t}(\ve{\defi{\sim}{k}}) - \hat{\Sigma}_{cl}(i\omega_n)  
\EEI
where $\defi{\sim}{\ve{k}}$ is the wave vector associated with translational invariance from cluster to cluster, $\mu$ is the chemical potential, $ \hat{t}(\ve{\defi{\sim}{k}})$ the full hopping matrix (including intra-cluster and inter-cluster hoppings) and $\hat{\Sigma}_{cl}(i\omega_n)$ is the self-energy of the cluster, imposed to be equal to the self-energy of the lattice in the CDMFT approximation. The hat on symbols specifies that they are matrices in the basis of cluster Wannier states. 

The self-energy is related to the Green function of the cluster through Dyson's equation 
\BEI
\label{eq:usualDysonCluster}
\hat{G}_{cl}^{-1}=\hat{\mathcal{G}}_{0}^{-1} -\hat{\Sigma}_{cl} 
\EEI
where the free propagator on the cluster is defined by
\BEI
\label{eq:DynamicalMeanField}
\hat{\mathcal{G}}_{0}^{-1}(i\omega_n) & = & \left( i\omega_n + \mu \right) \hat{\mathbb{I}} -\hat{h}^{0}_{loc}(i\omega_n) - \hat{\Delta}_{cl}(i\omega_n),
\EEI
with $\hat{h}^{0}_{loc}$ the one-body part of the Hamiltonian and $\hat{\Delta}_{cl}$ the hybridization function that defines the bath and its coupling to the cluster.  

%

The projection of the lattice Green function on the cluster leads to the self-consistent equation for the hybridization function
\BEI
\label{eq:ClusterProjection}
\hat{G}_{cl}(i\omega_n) & = & N_{cl}\int \frac{d\tilde{\ve{k}}}{(2\pi)^2}   \hat{G}_{latt}(i\omega_n,\defi{\sim}{\ve{k}}),
\EEI
where $N_{cl}$ is the number of sites on the cluster and the integral is over the reduced Brillouin zone. A high frequency expansion of both sides of this equation proves that $\hat{h}^{0}_{loc}$ is the hopping matrix within the cluster.~\cite{Koch_Sum_Rules_2008} 

The uniform spin susceptibility, or Knight shift, is defined  by
\begin{align}
	\NN 
	S_z &=\frac{1}{2} \left(N_{\uparrow}-N_{\downarrow}\right)\\
	\chi_{z}(q=0,\omega=0) &= \int_{0}^{\beta} \left\langle S_z(\tau) S_z(0) \right\rangle d\tau,
	\label{eq:Susceptibility}
\end{align}
with $N_{\uparrow}$ and $N_{\downarrow}$ the total number of up and down spins, respectively, on the cluster. 

The calculation starts from a guess for the hybridization function $\hat{\Delta}_{cl}$. This gives the so-called dynamical mean-field, Eq.~\ref{eq:DynamicalMeanField}, namely the free propagator for the cluster. Assuming that the impurity problem can be solved, Dyson's Eq.~\ref{eq:usualDysonCluster} then gives the self-energy which is needed to obtain the lattice Green function, Eq.~\ref{eq:GreenLattice}, entering the right-hand side of the self-consistency Eq.~\ref{eq:ClusterProjection}.  Using Dyson's equation again on the left-hand side of that self-consistency equation leads to a new guess for the hybridization function. This process is iterated until convergence. 

Calculating the cluster impurity Green function $\hat{G}_{cl}$ is the difficult problem. This is done here using  a continuous-time quantum Monte Carlo method (CTQMC).

\subsection{Hybridization-Expansion Continuous-Time Quantum Monte Carlo}

 CTQMC provides a statistically exact solution of the impurity problem exempt from imaginary-time discretization error. The large values of \UU,  low temperatures and large frustration that we need can be attained only with the  hybridization expansion algorithm (CT-HYB).~\cite{Gull_Performance_2007} Extensive reviews of CTQMC solvers are available.~\cite{Gull_Continuous-time_2011,Werner_CTQMC,Haule_CTQMC}
 
The cluster that tiles the infinite anisotropic triangular lattice is illustrated in Fig.~\ref{fig:AnisotropicSquareLattice}. It allows a singlet ground state. To speedup the calculations, one chooses a single-particle basis that transforms as the irreducible representations of the cluster-Hamiltonian symmetries.~\cite{Haule_CTQMC} The point group symmetry $C_{2v}$ of the anisotropic cluster as well as charge and spin conservation lead to the following single-particle basis
\BEI
\NN c_{A_{1} \sigma} & = & \frac{1}{\sqrt{2}} \left( c_{1 \sigma} + c_{3 \sigma} \right)
\\
\label{eq:Irreducible} c_{A_{1} \sigma}' & = & \frac{1}{\sqrt{2}} \left( c_{2 \sigma} + c_{4 \sigma} \right)
\\
\NN c_{B_{1} \sigma} & = & \frac{1}{\sqrt{2}} \left( c_{1 \sigma} - c_{3 \sigma} \right)
\\
\NN c_{B_{2} \sigma} & = & \frac{1}{\sqrt{2}} \left( c_{2 \sigma} - c_{4 \sigma} \right),
\EEI
where the indices are those of Fig. \ref{fig:AnisotropicSquareLattice} and where $A_{1}$, $B_{1}$, $B_{2}$ are irreducible representations of $C_{2v}$,  $A_{2}$ being empty. In this basis, the hybridization function $\hat{\Delta}$  and cluster Green function are both block diagonal. The largest block is $2\times2$ because the $A_{1}$ representation occurs twice. The calculations presented here are possible only if the angle defining rotations in this $2\times2$ block is chosen to minimize the sign problem.~\cite{Semon_Importance_2012}   
In addition it is necessary to use a modification of the original algorithms to ensure ergodicity in the presence of d-wave superconductivity.~\cite{Semon_Ergodicity_2014} We also speedup the calculation with the Lazy Skip List algorithm.~ \cite{Semon_Lazy_2014}

In normal phase studies, this basis respects the symmetries of the lattice that are compatible with the partitioning. In broken symmetry phases, such as magnetically ordered or d-wave SC, symmetry-breaking is allowed only for the hybridization function. The cluster continues to respect the original Hamiltonian symmetries. There is no mean-field factorization on the cluster. 

Other popular continuous-time quantum Monte Carlo impurity solvers involve expansion in powers of the interaction. They have better scaling than CT-HYB with cluster size. However, they need very large order expansion at large $U/t$, which makes them converge slowly, and they have a severe sign problem at large interaction strengths $U/t$ and frustration $t'/t$.~\cite{Gull_Continuous-time_2011} 

\subsection{Broken symmetry phases}

The Green function for superconductivity is written in Nambu notation as 

\BEI
\label{eq:GreenSupra}
-\langle\mathcal{T}_\tau \mathbf{\Psi}\mathbf{\Psi}^\dagger (\tau) \rangle  & = & \left( \begin{matrix}
	\hat{G}_{\uparrow}(\tau) && \hat{\mathcal{F}}(\tau)\\
	\hat{\mathcal{F}}^{\dagger}(\tau) && -\hat{G}_{\downarrow}(-\tau)
\end{matrix} \right),
\EEI
with $\mathbf{\Psi}^\dagger = (\mathbf{c}^\dagger_\uparrow, \mathbf{c}_\downarrow)$ where $\mathbf{c}^\dagger_\uparrow$ and $\mathbf{c}_\downarrow$ are row vectors as defined by equation Eq. \ref{eq:Irreducible}. The d-wave superconducting order parameter transforms as the $A_2$ representation of the $C_{2v}$ symmetry group. Hence only entries in the Gork'ov function $\hat{\mathcal{F}}(\tau)$ transforming as $A_2$ can be finite, e.g. for singlet pairing 
\begin{align}
\begin{split}
\mathcal{F}_{B_1,B_2}(\tau) :&= -<\mathcal{T}_\tau c_{B_1\uparrow}(\tau) c_{B_2\downarrow}> \\
&= -<\mathcal{T}_\tau c_{B_2\uparrow}(-\tau)c_{B_1\downarrow}> \\
&=: \mathcal{F}_{B_2,B_1}(-\tau)
\end{split}
\end{align}

To determine the region where the SC phase is allowed we calculate the order parameter
\BEI
\label{eq:ParametreSupra}
d_{SC} := \mathcal{F}_{B_1,B_2}(0^+).
\EEI

%
%
%
%

Since antiferromagnetism does not break $C_{2v}$ symmetry, no additional entry is needed in the Green function matrix. We only need to let up and down spins take independent values. 

\section{A brief review of quantum cluster results for the cuprates}\label{Sec:cuprates}

In this section we only briefly summarize some of the main results obtained with cluster generalizations of DMFT for the cuprates and give a few representative references. A more detailed but not exhaustive review can be found in Ref.~\onlinecite{TremblayJulichPavarini:2013}. 

There are two cluster generalizations~\cite{Maier_Quantum_2005,KotliarRMP:2006,LTP:2006} of DMFT. We described CDMFT above. In the Dynamical Cluster Approximation (DCA)~\cite{Hettler:1998} the clusters are built in momentum space. Whatever the method used, all groups have found a pseudogap in the normal state near half-filling.~\cite{Huscroft:2001,Senechal:2004,kyung:2006b,LiebschTong:2009,Lin:2009,Ferrero:2009,Sordi_Widom} The Mott transition at half-filling is first-order,~\cite{park:2008,Balzer:2009} and large cluster studies find~\cite{Maier_Cluster_Size_SC} a d-wave superconducting transition temperature $\TC$ at finite doping. The zero-temperature order parameter has a dome shape instead of increasing monotonously towards half-filling when the interaction strength is large enough that there is a Mott insulator at half-filling or when antiferromagnetism is allowed to compete with superconductivity.~\cite{Senechal:2005,CaponeKotliarAFM_SC:2006,Haule:2007,Aichhorn:2006,Kancharla:2008,Gull:2013,Frattini:2015} The larger cluster studies~\cite{Gull:2013} find that the dome ends at a finite doping away from half-filling. 

All methods agree with the existence of crossovers at high-temperature associated with the opening of a pseudogap as doping is reduced towards half-filling. On $2\times2$ clusters with the CT-HYB solver it was possible to reach lower temperatures than previous studies. Scanning chemical potential over a very fine mesh allowed the discovery of a first-order transition in the normal state at finite doping.~\cite{Sordi_Widom} That first-order transition ends at a critical point that is continued as a Widom line in the supercritical region. That first-order transition with its Widom line then becomes an organizing principle for the observed crossovers and for the superconducting dome.  The Widom line is described in more details in the following section that begins our discussion of organics. Earlier work on normal state first-order transitions at finite-doping is discussed at the end of that section.

\section{Normal state pseudogap, first-order transition, and Widom line}\label{Sec:normal}

The first-order Mott transition at half-filling is well documented.~\cite{Liebsch_Mott_Triangle_2009,Ohashi_Finite_2008} The blue shaded region in Fig.~\ref{fig:PlaquetteTp04-n100} identifies the region of parameter space where normal-state hysteresis is found. Metallic and insulating states there can coexist.  The results are similar to those obtained at half-filling in the unfrustrated case $t'=0$.~\cite{park:2008,Balzer:2009} The positive slope of the transition in the $T-U$ plane (negative in the $T-t/U$ plane) comes from the smaller entropy of the insulating phase compared with the metallic phase, as deduced from Clausius-Clapeyron arguments.~\cite{Lefebvre_Phase_Diagram_2000,Sordi_Metal_Transitions_2011} Indeed, from $dG=-SdT+DdU+\mu dn$, where $G$ is the Gibbs free energy, $S$ the entropy, $D$ double occupancy, and $n$ the filling, we find that the slope of the transition line is
\begin{equation}
\frac{dT}{dU}=\frac{D_I-D_M}{S_I-S_M}
\end{equation} 
where index $I$ is for the insulating phase and $M$ for the metallic one. The smaller entropy of the insulating phase comes from the tendency to form local singlets.~\cite{Lefebvre_Phase_Diagram_2000,Sordi_Finite_Doping_2010,Sordi_Metal_Transitions_2011,Sordi_Widom}

\begin{figure}
	\includegraphics[width=0.95\linewidth]{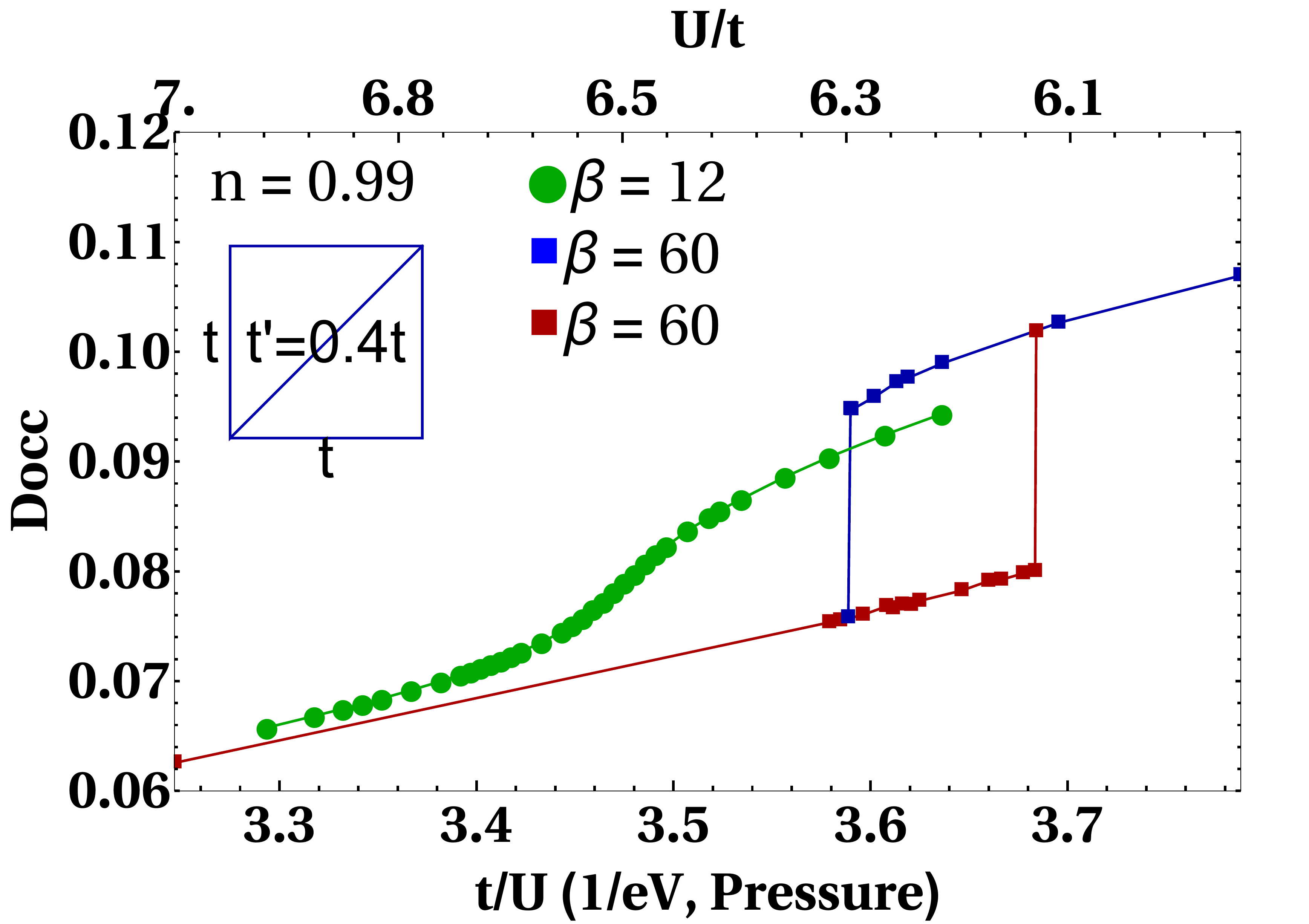}	
	\caption{Double occupancy $D_{occ}$ as a function of pressure (bottom horizontal axis) or interaction strength $U/t$ (top horizontal axis) for fixed filling, $n=0.99$. The value  $t=0.044$ eV is used to convert to physical units.~\cite{Liebsch_Mott_Triangle_2009} The lower horizontal axis is labeled $t/U$ to suggest the pressure dependence, but the numbers on that horizontal axis are given by the value of $1/U$  expressed in electron-Volt using the above conversion factor. At $T=t/60$ there is a first-order hysteresis region: the brown squares are obtained from the insulating solution and the blue squares for the conducting solution. At $T=t/12$, there is no hysteresis, only an inflexion point that determines the Widom line.}
	\label{Fig:transition_99_D(U)}
\end{figure}

In this section, we focus on the less familiar first-order transition found at finite doping. Consider the case $t'=0.4t$ at $1\%$ doping, namely $n=0.99$. Fig.~\ref{Fig:transition_99_D(U)} shows two jumps of double-occupancy delimiting a coexistence region at low temperature, and a smooth dependence on \UU at high temperature. The jumps at low-temperature define the hysteresis region of a first-order transition. The low pressure phase exhibits a pseudogap while the high-pressure phase is a more standard metal. The inflexion point at high temperature defines a crossover. The locus of these inflexion points is associated with the so-called Widom line of the first-order transition. 

\begin{figure}
	\includegraphics[width=0.95\linewidth]{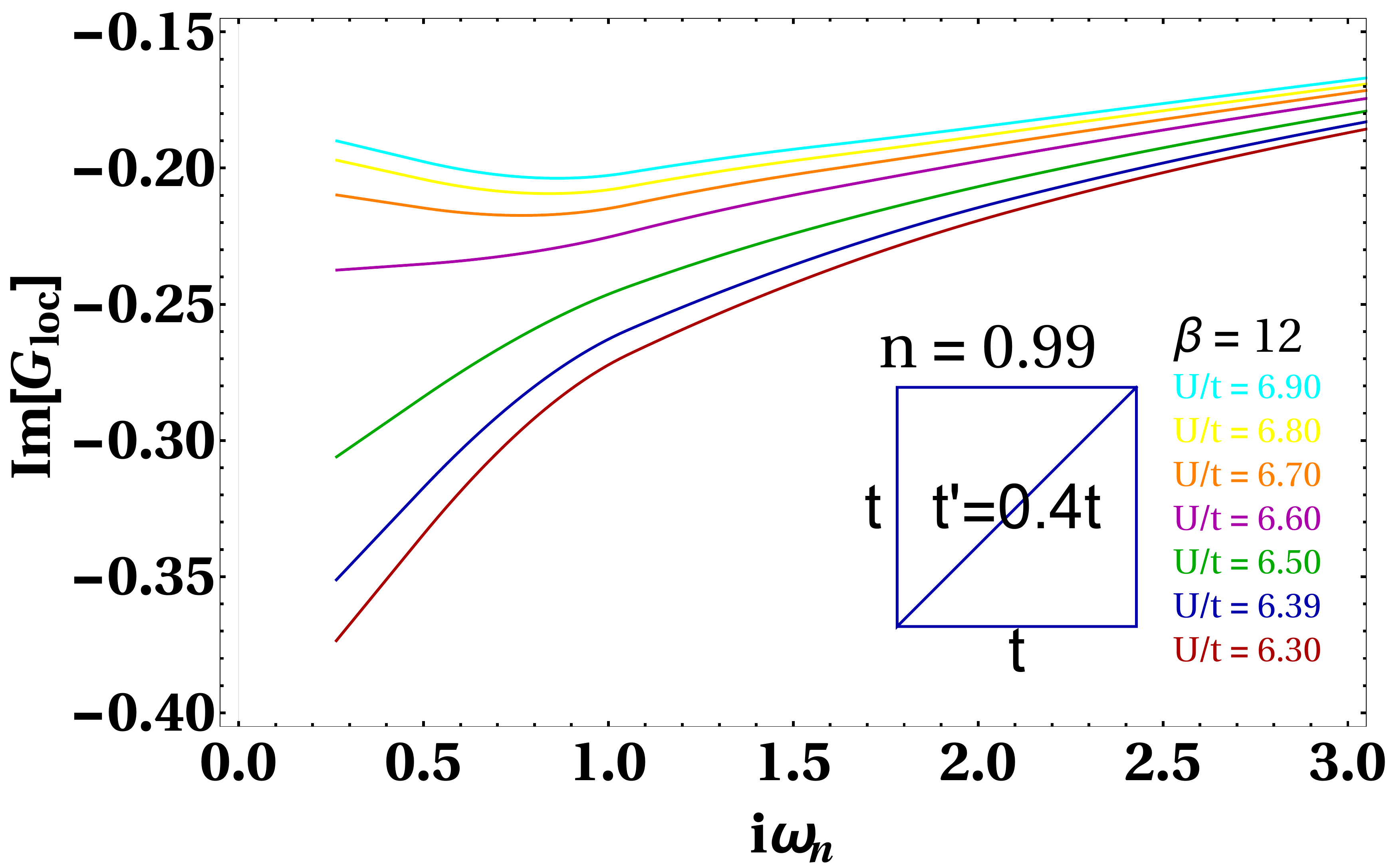}	
	\caption{The imaginary part of the Matsubara Green function $\operatorname{Im} G(i\omega_n)$ plotted as a function of Matsubara frequency gives information about the density of states at the Fermi level, $-2\operatorname{Im} G(i\omega_n\rightarrow i\eta)$ with $\eta\rightarrow 0$. The behavior differs depending on the value of \UU. The decrease towards zero or the density of states at larger \UU indicates a pseudogap, while its increase at smaller \UU indicates a metallic phase.}
	\label{Fig:transition_99_ImG}
\end{figure}

In the theory of fluids, the Widom line is defined as the line where the maxima of different thermodynamic response functions touch each other asymptotically as one approaches the critical point of the first-order transition.~\cite{XuStanleyWidom:2005} Investigations on the phase diagram of fluids have shown drastic changes in the dynamics upon crossing the Widom line. ~\cite{XuStanleyWidom:2005,Simeoni2010} By analogy, in the cuprates the Widom line has been identified as the organizing principle for the pseudogap and resulting phase diagram of the high-temperature superconductors.~\cite{Sordi_Widom,Sordi_c_axis_2013,Alloul_Widom:2013,Taillefer:2015} 

In the present context, along the Widom line a crossover from a metallic state to a pseudogap metal  is also seen. This is illustrated by the frequency dependence of the imaginary part of the {\it local} Matsubara Green function $\operatorname{Im}(G(i\omega_n))$ in Fig.~\ref{Fig:transition_99_ImG}  at $T$ slightly above the critical point of the first-order transition. For large values of \UU,  $\operatorname{Im}(G(i\omega_n))$ aims upwards as frequency decreases, indicating a small density of states at the Fermi level, consistent with a pseudogap. By contrast, for smaller values of \UU, $\operatorname{Im}(G(i\omega_n))$ extrapolates to a finite density of states at the Fermi level, consistent with an ordinary metal. We suggest that this crossover corresponds to the one seen experimentally in doped organics.~\cite{taniguchi_anomalous_2007,Oike_Mottness_2014} At lower temperature, the transition between the pseudogap metal and the more ordinary metallic phase occurs discontinuously through the first-order transition illustrated in Fig.~\ref{Fig:transition_99_D(U)}. 

Fig.~\ref{fig:PlaquetteTp04-n099} displays the normal state phase diagram at $1\%$ doping. There is a coexistence region, in blue, coming from the first-order transition and a Widom line that extends above the critical point of that first-order transition. There is also a Widom line in the half-filled case (not shown). In the context of the cuprates, this first-order transition was found at fixed \UU  as a function of doping.~\cite{Sordi_Finite_Doping_2010,Sordi_Metal_Transitions_2011,Sordi_Widom}  The results in those papers~\cite{Sordi_Finite_Doping_2010,Sordi_Metal_Transitions_2011,Sordi_Widom} clearly show a surface of first-order transitions that is continuously connected to the Mott transition at half-filling.~\footnote{The critical line ending the first-order surface plunges to very low temperature as $U$ or doping increase. The first-order transition may end at large $U$ or continue as a quantum-critical line. We can only confirm that crossovers exist at high temperature. The sign problem prevents us from exploring very low temperature.} The results of Fig.~\ref{fig:PlaquetteTp04-n099} are in a way a constant doping cut of the finite $t'$ version of that first-order surface. The critical point in Fig.~\ref{fig:PlaquetteTp04-n099}  occurs at a temperature about 60\% lower than the corresponding temperature at $n=1 $. That rapid drop is also observed in the square-lattice results.  

We end this section with brief comments on early work on the doping-induced Mott transition of the Hubbard model. In single-site DMFT,~\cite{Kotliar:2002} it was found for $t'=0$ that upon doping there is a first order transition between a half-filled Mott insulator and a finite-doping metal. Essentially the same result was found with DCA for the square lattice and various positive $t'/t$ (electron-doped case in the language of cuprates).~\cite{MacridinPS:2006} Later work with the same methods \cite{Khatami_first_order:2010} suggested that at $t'=0$ there is a quantum-critical point instead of a first-order transition but the lowest temperature reached was large compared to those where the first-order transition was found in CDMFT with CTQMC solver.~\cite{Sordi_Finite_Doping_2010,Sordi_Metal_Transitions_2011,Sordi_Widom} CDMFT studies with an exact diagonalization solver~\cite{LiebschTong:2009} also found a clear first-order transition for positive $t'/t$ on the cuprate square lattice. As in Refs.~~\cite{Sordi_Finite_Doping_2010,Sordi_Metal_Transitions_2011,Sordi_Widom}, that first-order transition separates a pseudogap phase and a metal instead of a separating a Mott insulator and a metal as found in the above early DCA study.~\cite{MacridinPS:2006}  In that study,~\cite{MacridinPS:2006} it was noted that the phase transition appeared only for $U/t$ larger than the bandwith. However, it was in the work of Refs.~\onlinecite{Sordi_Finite_Doping_2010,Sordi_Metal_Transitions_2011,Sordi_Widom} that the critical end line of the finite-doping surface of first-order transitions in the $(U,T,\delta)$ space of parameters was shown to be connected to the critical end point of the Mott transition at half-filling. This is an important step to differentiate strong and weak correlation effects.~\cite{TremblayJulichPavarini:2013} The pseudogap phase appears at finite doping only if there is a Mott insulator at half-filling. At zero temperature the transition between the Mott insulator and the pseudogap phase is second order.~\cite{Sordi_Finite_Doping_2010,Sordi_Metal_Transitions_2011} The pseudogap is then different from the Mott gap even though they both appear in the generalized phase diagram.~\cite{Sordi_Widom} The significance and existence of the Widom line was noted in Refs.~\onlinecite{Sordi_Widom,Sordi_c_axis_2013,Alloul_Widom:2013}.

\begin{figure*}[!htbp]
	\subfigure[]
	{
		\includegraphics[width=0.45\linewidth]{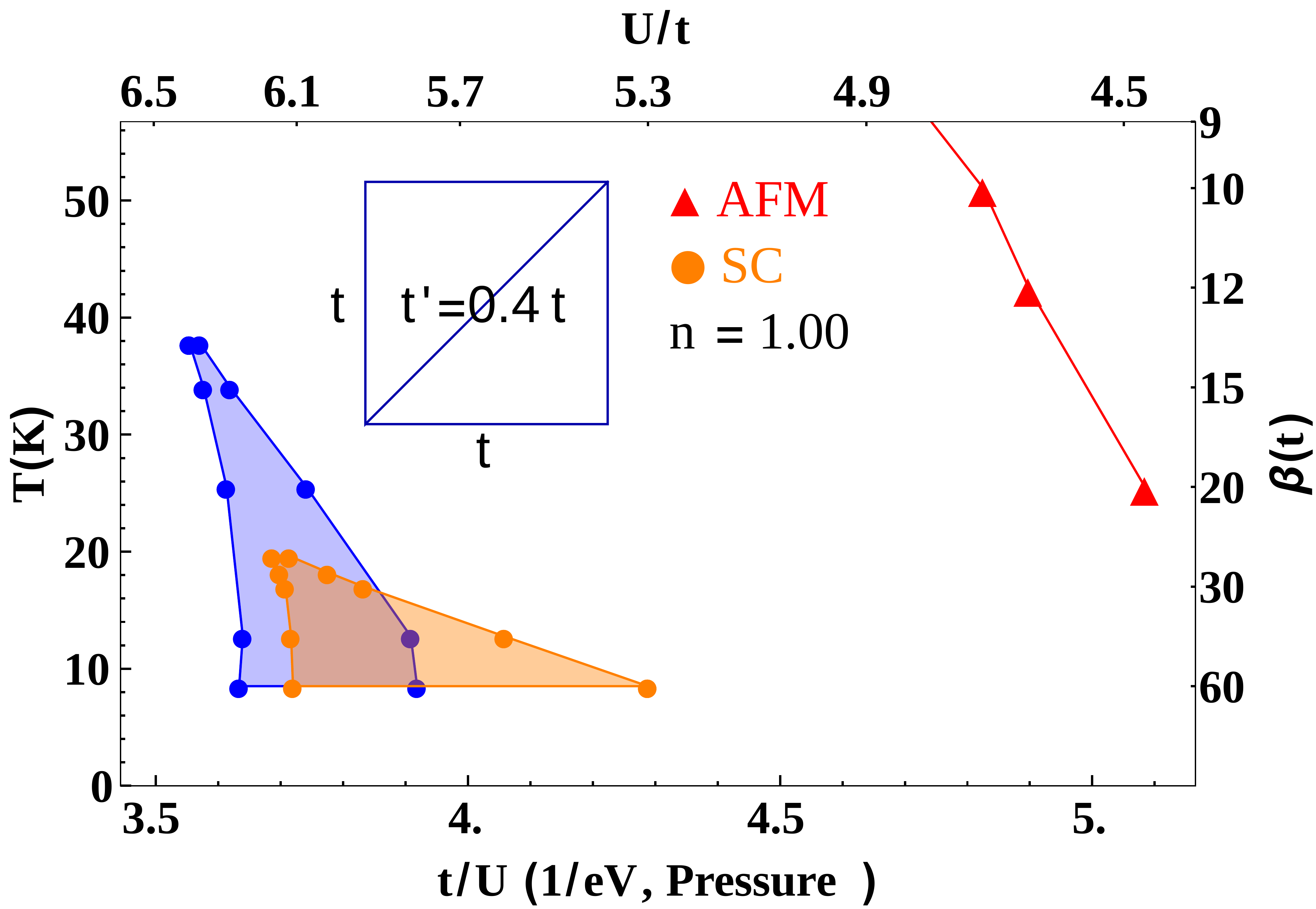}
		\label{fig:PlaquetteTp04-n100}
	}
	\subfigure[]
	{
		\includegraphics[width=0.45\linewidth]{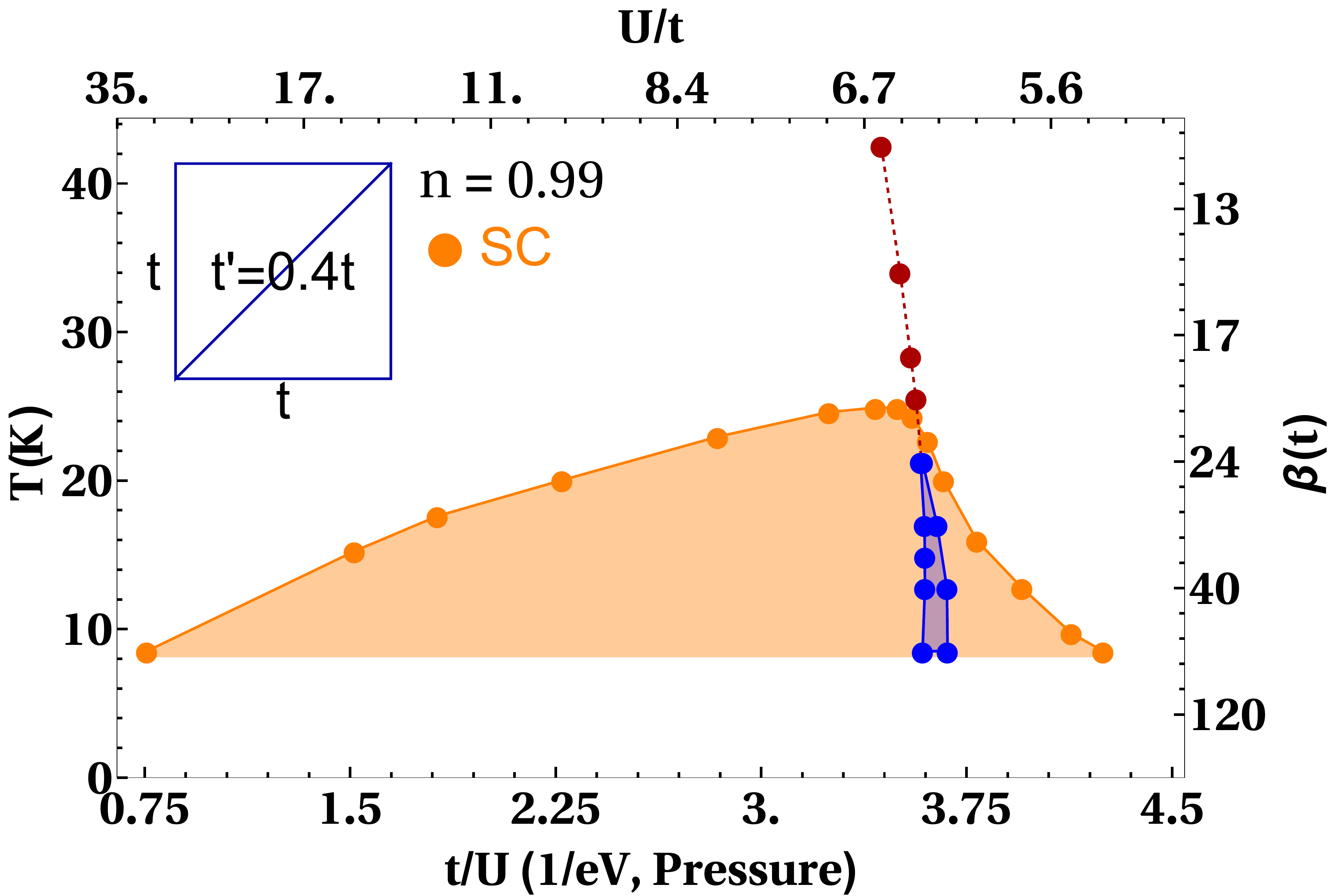}
		\label{fig:PlaquetteTp04-n099}
	}
	\newline
	\subfigure[]
	{\includegraphics[width=0.45\linewidth]{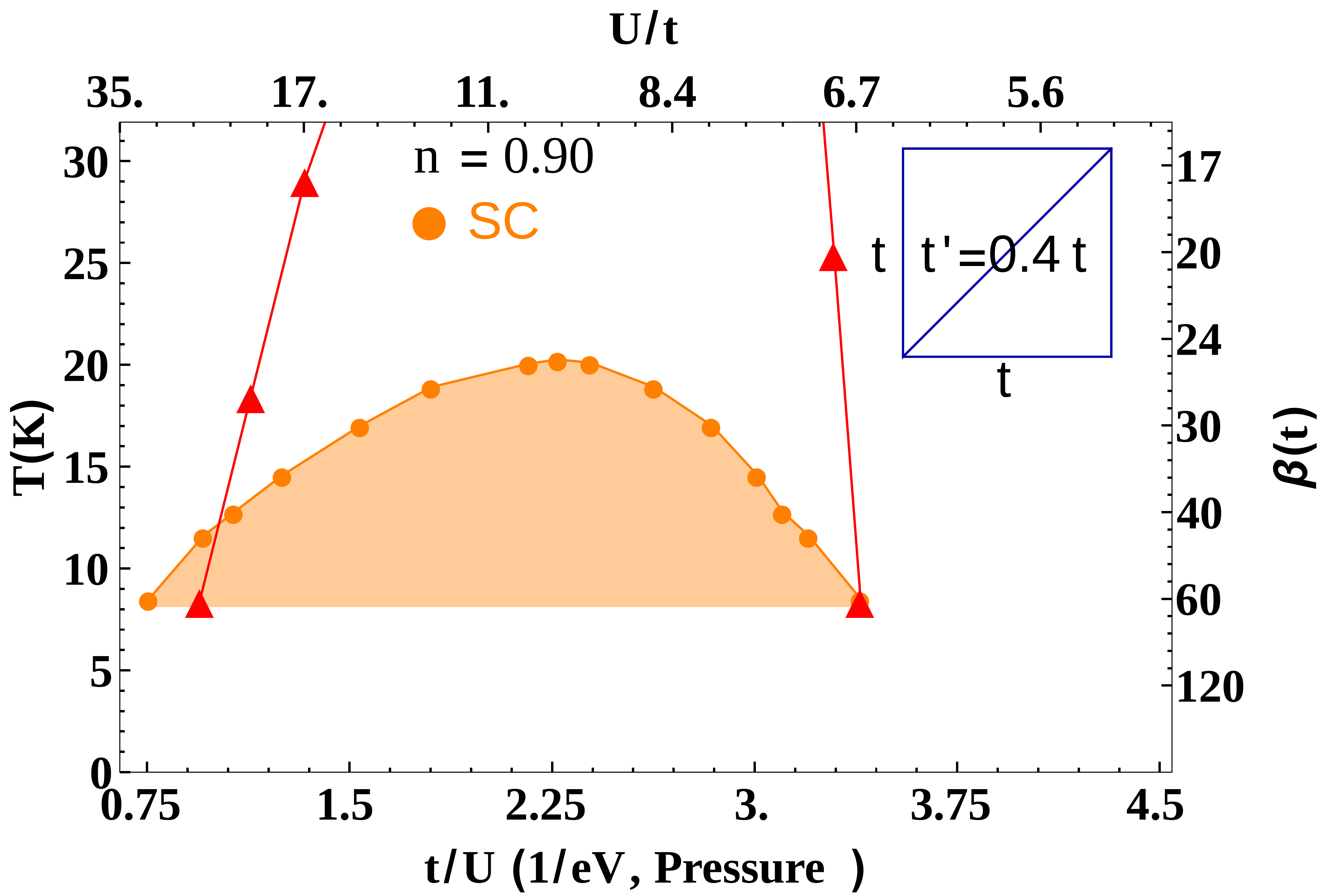}
		\label{fig:PlaquetteTp04-n090}
	}
	\subfigure[]
	{\includegraphics[width=0.45\linewidth]{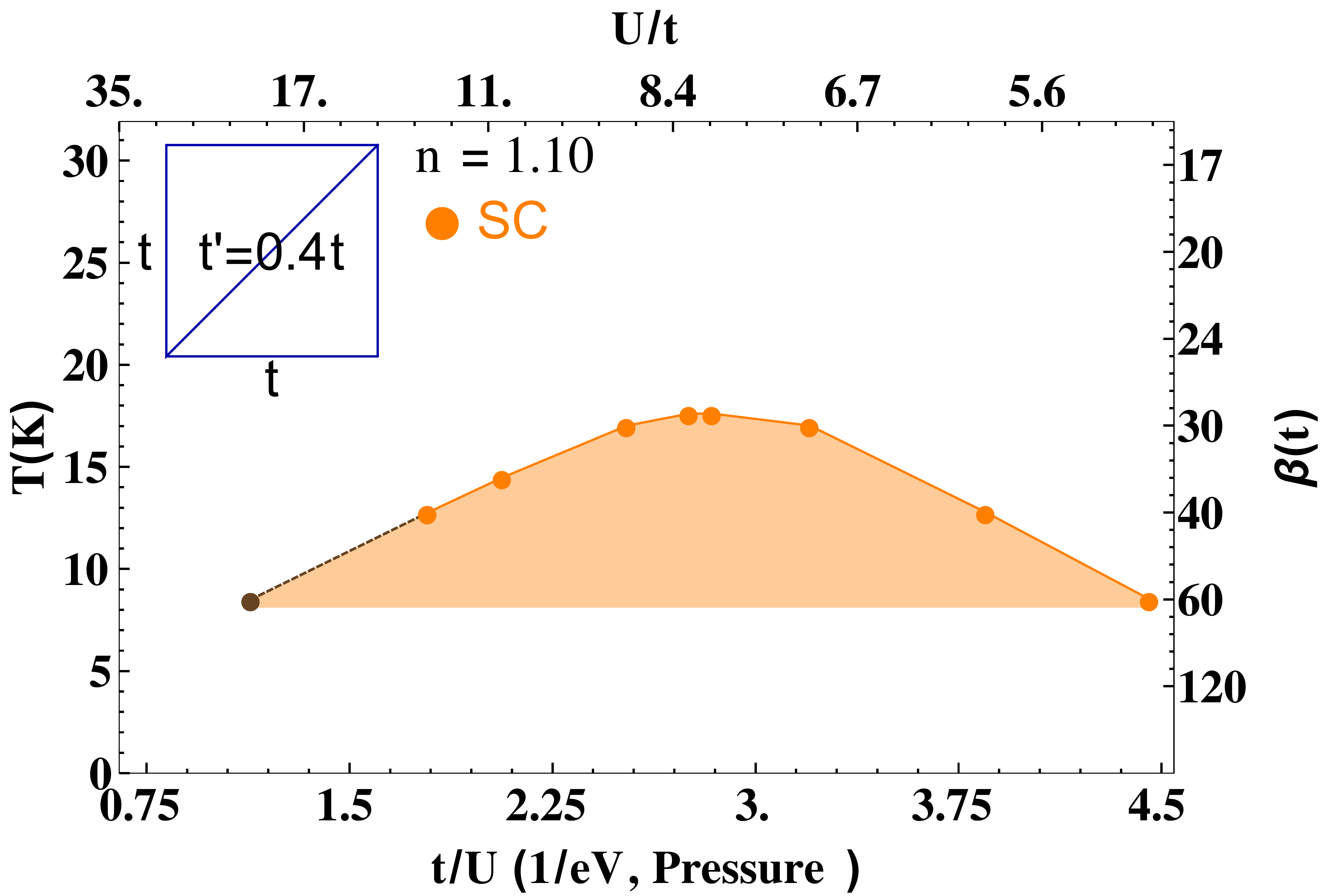}
		\label{fig:PlaquetteTp04-n110}
	}
	\caption{
		(color online) Phase diagrams for the Hubbard model on the anisotropic triangular lattice  with  $t'=0.4t$ for various fillings. The antiferromagnetic state has been studied for the half-filled case and for $10\%$ doping. The value  $t=0.044$ eV is used to convert to physical units.~\cite{Liebsch_Mott_Triangle_2009} The lower horizontal axis is labeled $t/U$ to suggest the pressure dependence, but the numbers on that horizontal axis are given by the value of $1/U$  expressed in electron-Volt using the above conversion factor. The same convention is used throughout the paper. Lines are guides to the eye. 
		\newline
		\tg{(a)} Phase diagram for the half-filled case. In the blue region, the Mott insulator and the metallic state coexist. At the blue points, we find a first-order jump in the normal state double occupancy. The lines between the points can be identified as the spinodal lines. d-wave superconductivity occurs in the orange region: Decreasing pressure, we find a first-order jump of the SC order parameter to zero at the orange points while upon increasing pressure we find a second-order transition. The AFM phase occurs below the red line that interpolates between the triangles where we find the N\'{e}el second-order transition. 
		\newline
		\tg{(b)} Phase diagram for the $1\%$ hole-doped case. The colors have the same meaning as for the half-filled case, except that in the blue region two different metallic states are found instead of a metallic and an insulating state like at half-filling. First-order jumps are observed at the blue points. There is coexistence in the blue region. Also, the red dots connected to the blue region indicate a strong crossover (Widom line) between a pseudogap state at small pressure (large \UU) and a metallic state at large pressure (small \UU).  Orange points denote where we detect a second-order transition from the SC state to the normal state.
		\newline
		\tg{(c)}  Phase diagram for the $10\%$ hole-doped case. The transitions between the normal and SC state (orange circles) are second-order. The AFM phase is between the red lines. These lines interpolate between the second-order N\'eel transition that we find where the triangles are located.
		\newline
		\tg{(d)} Phase diagram for the $10\%$ electron-doped Hubbard model. The brown dot and line are extrapolations. The transitions between the normal and SC state (orange circles) are second-order. 
	}
	\label{fig:PlaquetteDopageDiagramtp04}
\end{figure*}

\section{Phase diagram, including superconductivity and antiferromagnetism}\label{sec:results}

We present phase diagrams for the normal state, N\'{e}el antiferromagnetism (AFM) and  $d$-wave superconductivity (SC).  In CDMFT, AFM and SC fluctuations are treated on equal footing both on the cluster and in the bath but we only allowed one symmetry-breaking at a time. Values for $t'/t$ are inspired from Kandpal and \textit{al.} \cite{Kandpal_Revision_Parameters_2009} and from Nakamura and  \textit{al.}~\cite{Nakamura_Revision_Parameters_2009} Using \textit{ab initio} density functional theory, they found  that \KCl could be modeled by $t'=0.4t$ and \KCN by $t'=0.8t$. However, extensive 
H$\defi{..}{\text{u}}$ckel 
calculations had previously found higher frustration ($t^\prime/t$) values for these compounds~\cite{Huckel}. We have to keep in  mind these uncertainties when we make contact with real materials.

We present first the case $t'=0.4t$, then $t'=0.8t$.  The intermediate frustration $t'=0.4t$ results are presented first because the sign problem is less severe in that case, allowing a more thorough investigation of the SC phase diagram. Even though the SC phase would be mostly hidden by antiferromagnetism in this case, we find that the results for the pure SC phase (forbidding AFM) are qualitatively similar to the results we discuss for larger frustration $t'=0.8t$. In the later case, AFM is generally sufficiently suppressed that its neglect is justified. We are restricted to commensurate antiferromagnetism. We also did not allow non-collinear spin order. A more in-depth discussion of magnetic order and its impact on the presence of  SC in real compounds can be found in  subsections~\ref{subsec:Tp0.8} and ~\ref{sec:discussion_broken_symmetry}. 


\subsection{$t' = 0.4t$}
\label{subsec:Tp0.4}

\paragraph*{} 
Our results for the phase diagrams at different dopings are summarized in Fig.~\ref{fig:PlaquetteDopageDiagramtp04}. A few more properties for the SC phase are displayed in Fig.~\ref{fig:ChiSz-OrderParametertp0.4}. Let us discuss the various phases in turn. 

\paragraph*{The phase diagram at half-filling} 
Fig.~\ref{fig:PlaquetteTp04-n100} depicts the phase diagram at half-filling. The blue region delimits the metal-insulator coexistence region associated with the first-order Mott transition in the normal phase. The critical value of \UU agrees within error bars with previous results.~\cite{Kyung_Mott_2006,Clay:2008} The d-wave SC phase is observed in proximity to the normal-state first-order  Mott transition.  When pressure is increased in the SC phase, (interaction strength is decrease), it disappears in a second-order manner. The zero-temperature results obtained previously~\cite{Kyung_Mott_2006} suggest that in that limit the SC phase in Fig.~\ref{fig:PlaquetteTp04-n100} will extend beyond the phase boundary for AFM. When pressure is decreased, SC gives way to the insulating phase through a first-order jump. The maximum SC critical temperature (\TCM) is attained close to the Mott transition to the insulator. All the qualitative results agree with previous theoretical studies on the unfrustrated square lattice at finite temperature \cite{Sordi_Strong_2012} as well as with experimental observations in various organic compounds of the \BEDT family and of the \DMITS family \cite{Lefebvre_Phase_Diagram_2000, Shimizu_Pressure_Induced_2010,Shimizu_Dmit}. 

Note that the low-pressure boundary (spinodal line) where the metastable metallic phase disappears discontinuously in favor of a stable insulator does not coincide with the low pressure boundary where the SC phase disappears discontinuously. The two boundaries are however in very  close proximity. There is no reason for the regions of metastability of the normal and superconducting phases to exactly coincide.  

When AFM order is permitted, it dominates a wide area in the temperature-pressure ($T-P$) plane. The maximum of the SC dome does not coincide with an AFM quantum critical point, as can be seen for example in~\ref{fig:PlaquetteTp04-n090}. Further results on AFM appear in Fig.~\ref{fig:PlaquetteHalfFillingTp0.4_AFM_Total} and are discussed in Secs.~\ref{subsec:Tp0.8} and ~\ref{sec:discussion_broken_symmetry}. Zero-temperature studies obtained with CDMFT~\cite{Kyung_Mott_2006} and with other methods~\cite{Watanabe:2008,Tocchio:2013} do suggest that for this value of $t'/t$ AFM is the most stable magnetic phase. 

\paragraph*{Phase transition between pseudogap phase and metallic phase} 
For small hole doping (1\% in Fig.~\ref{fig:PlaquetteTp04-n099}) one finds the first-order transition discussed in Sec.~\ref{Sec:normal}. The region where hysteresis is found is indicated by the blue region in Fig.~\ref{fig:PlaquetteTp04-n099}. Comparison with half-filling in Fig.~\ref{fig:PlaquetteTp04-n100} demonstrates that this transition is continuously connected to the first-order metal-insulator Mott transition and that it occurs at larger interaction strength as doping is increased, as suggested by the border of the magenta region in Fig.~\ref{fig:GeneralizedPhaseDiagram}. In other words, upon doping, the Mott insulator evolves continuously into a conducting pseudogap phase different from the metallic phase at larger pressure. This is in complete analogy with the results found previously~\cite{Sordi_Finite_Doping_2010,Sordi_Metal_Transitions_2011} for the unfrustrated square lattice. \footnote{On the square lattice,~\cite{Sordi_Finite_Doping_2010,Sordi_Metal_Transitions_2011} the chemical potential was varied on a fine scale whereas here it is the interaction that is varied slowly. The first-order transition occurs on a surface in the $T-U-\mu$ plane that can be traversed in several directions.} Our results are qualitatively similar to those of Fig.~1 of Ref.~\onlinecite{Sordi_Finite_Doping_2010} for the square lattice. We indeed also find that the first-order transition  occurs at lower temperatures as doping is increased and is not accessible to our simulations for hole-dopings as small as $4\%$ for temperatures down to $T/t=1/60$ and interaction strength up to \UU = 20. 

The type of pseudogap discussed here is a strong-correlation effect as follows from the fact that it appears in a phase that exists only for values of $U/t$ large enough for a Mott insulator to exist at half-filling. This pseudogap is however very different from the Mott gap.~\cite{Senechal:2004,Sordi_Widom,TremblayJulichPavarini:2013} 

\paragraph*{AFM} 
We did not systematically study the effect of doping on the AFM  phase. Nevertheless, Table~\ref{table:AFM} is suggestive. Hole-doping pushes the AFM phase to lower pressures ( higher interaction strengths \UU). For example, at $10\%$ hole-doping, the critical pressure for the N\'{e}el transition at $T/t=1/20$ is decreased by about 35\% (Fig.~\ref{fig:PlaquetteTp04-n090}) compared to the half-filled case. 
By contrast, electron-doping brings the N\'{e}el transition to higher pressure values (to lower interaction strengths \UU) (not shown here). A calculation of the Lindhard function shows that this can in part be attributed to better nesting in the electron-doped case. 

\begin{table}[htpb]
\begin{tabular}{|c|c|c|c|c|}
\hline \tg{n} & \tg{1.10} & \tg{1.00}  & \tg{0.99} & \tg{0.90}  \\ 
\hline \tg{U/t} &  $3.20 \pm 0.05$ & $4.465 \pm 0.005$  & $4.60 \pm 0.05$ & $6.9 \pm 0.1$ \\ 
\hline 
\end{tabular} 
\caption{Filling dependence of $U_{N}/t$ for $T/t = 1/20$. For $U/t > U_{N}/t$ the AFM phase is stable at that temperature} 
\label{table:AFM}
\end{table}

\begin{figure}[h]
\includegraphics[width=0.95\linewidth]{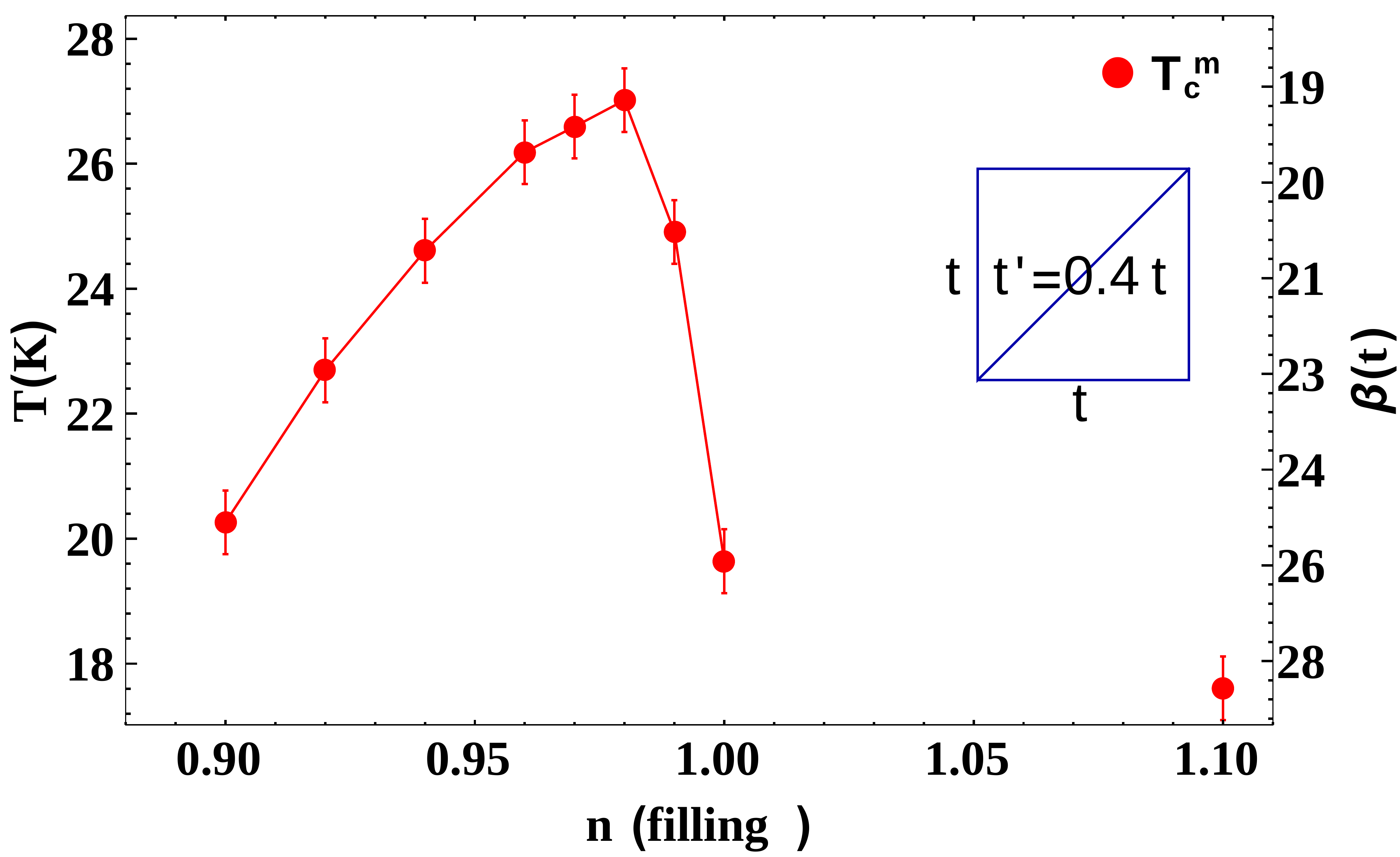}
\caption{Superconducting \TCM as a function of filling and \UU projected in the $T-n$ plane. The actual values of $T,U,n$ are listed in Table~\ref{table:SC}.
	}
	\label{fig:TcMaximum}
\end{figure}

\begin{figure}[]
	\includegraphics[width=0.95\linewidth]{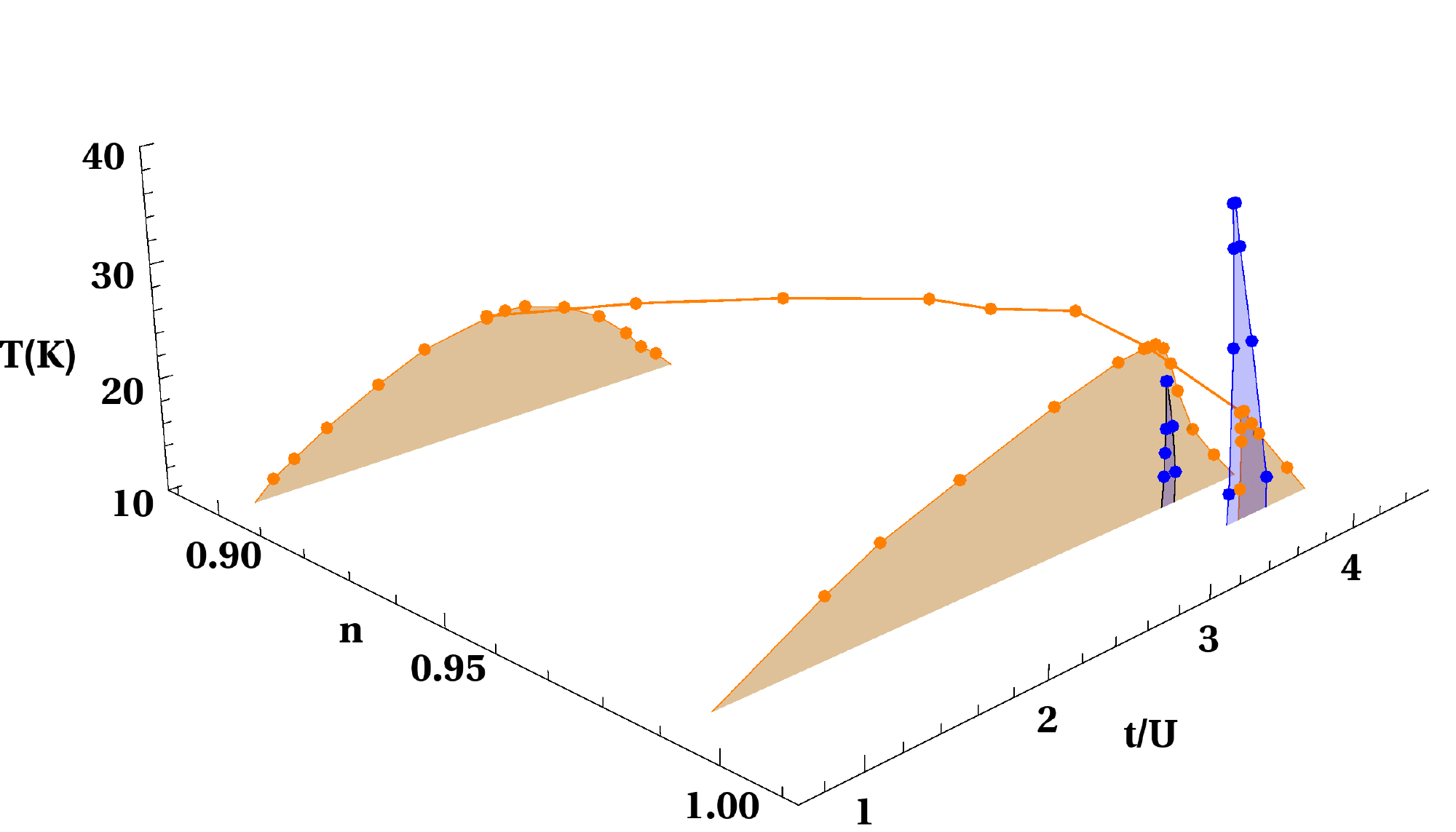}
	\caption{Superconducting phase diagram combining in $T-U-n$ space the constant-doping $T-U$ planes of Fig.~\ref{fig:PlaquetteDopageDiagramtp04}. The line of \TCM in Fig.~\ref{fig:TcMaximum} also appears on this plot.
	}
	\label{fig:3D}
\end{figure}

\begin{figure*}[!htbp]
	\centering
	
	\subfigure[]{
		\includegraphics[width=0.45\linewidth]{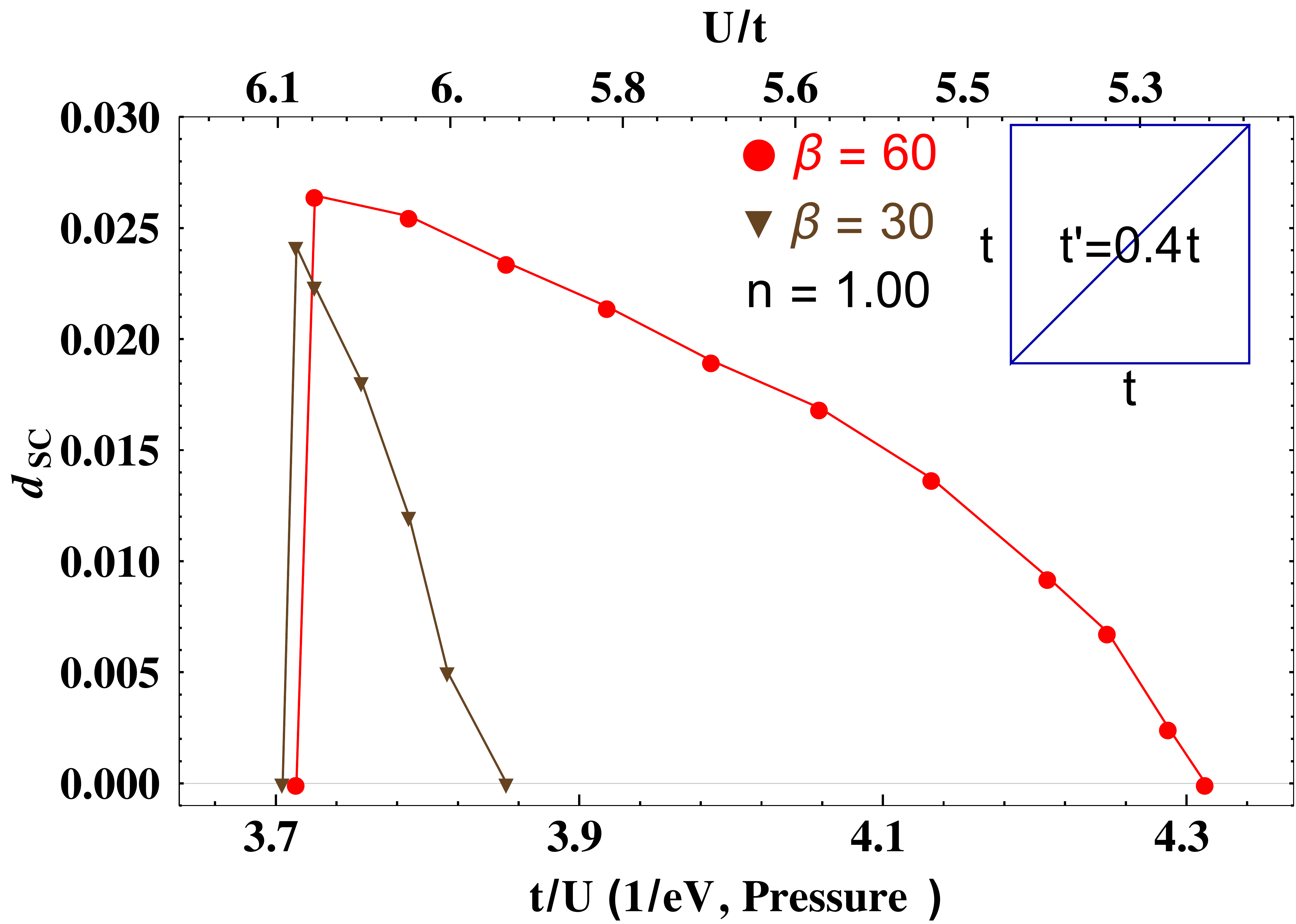}
		\label{fig:OrderParametertp0.4-Half-filling}
	}
	\subfigure[]{
		\includegraphics[width=0.45\linewidth]{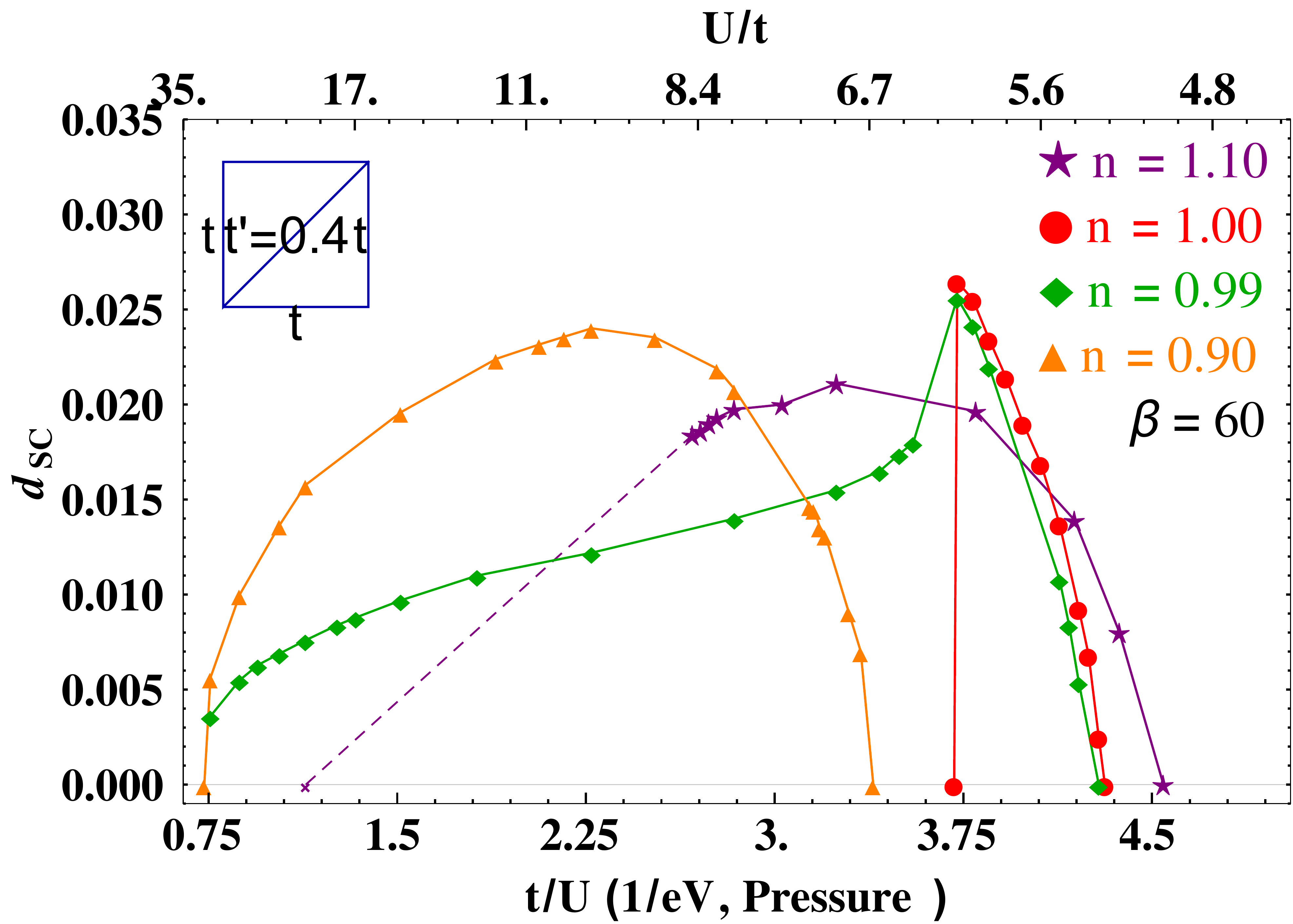}
		\label{fig:OrderParametertp0.4-All}
	}
	\newline
	\subfigure[]{
		\includegraphics[width=0.45\linewidth]{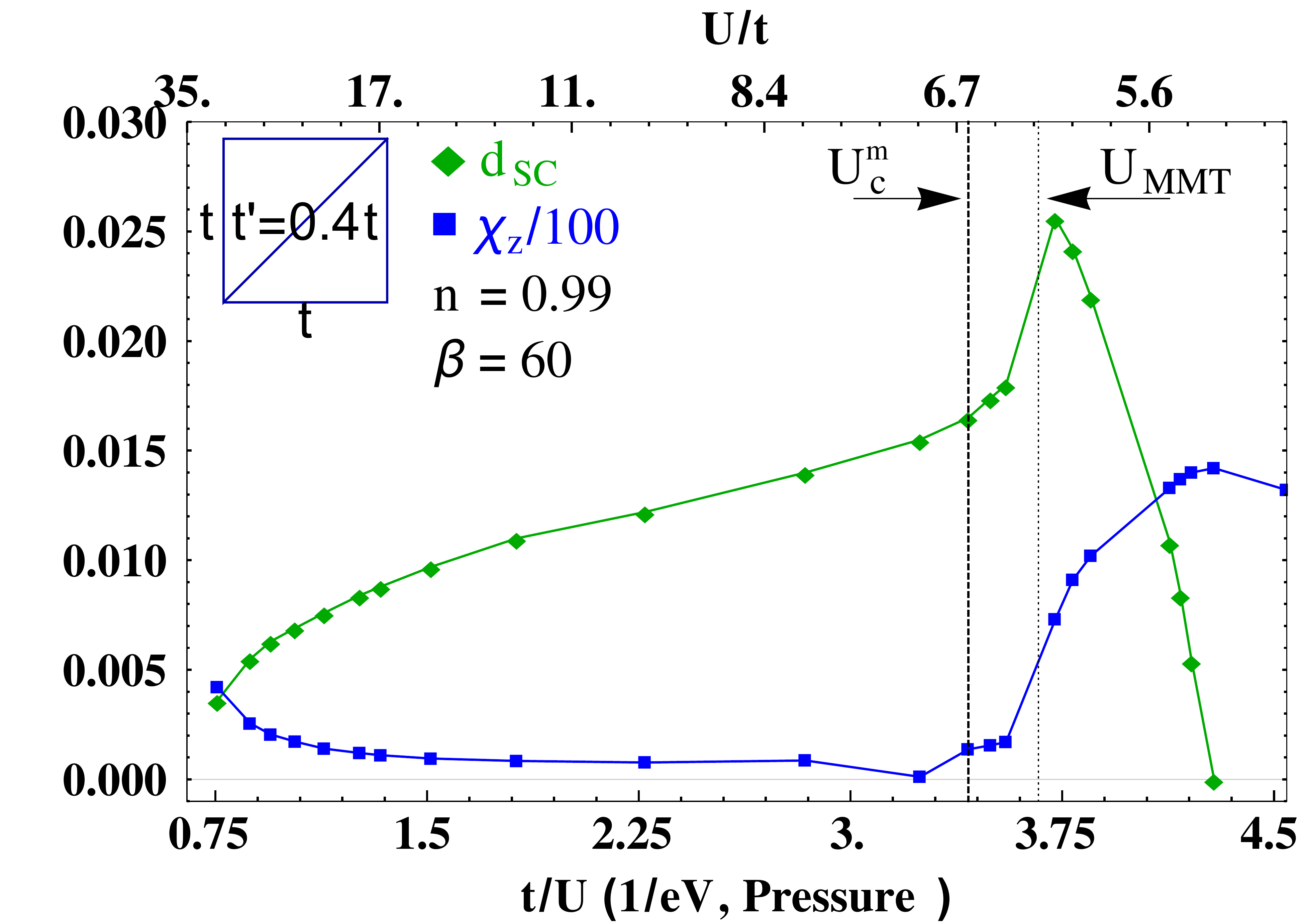}
		\label{fig:ChiSz-OrderParametertp0.4-n099}
	}
	\subfigure[]{
		\includegraphics[width=0.45\linewidth]{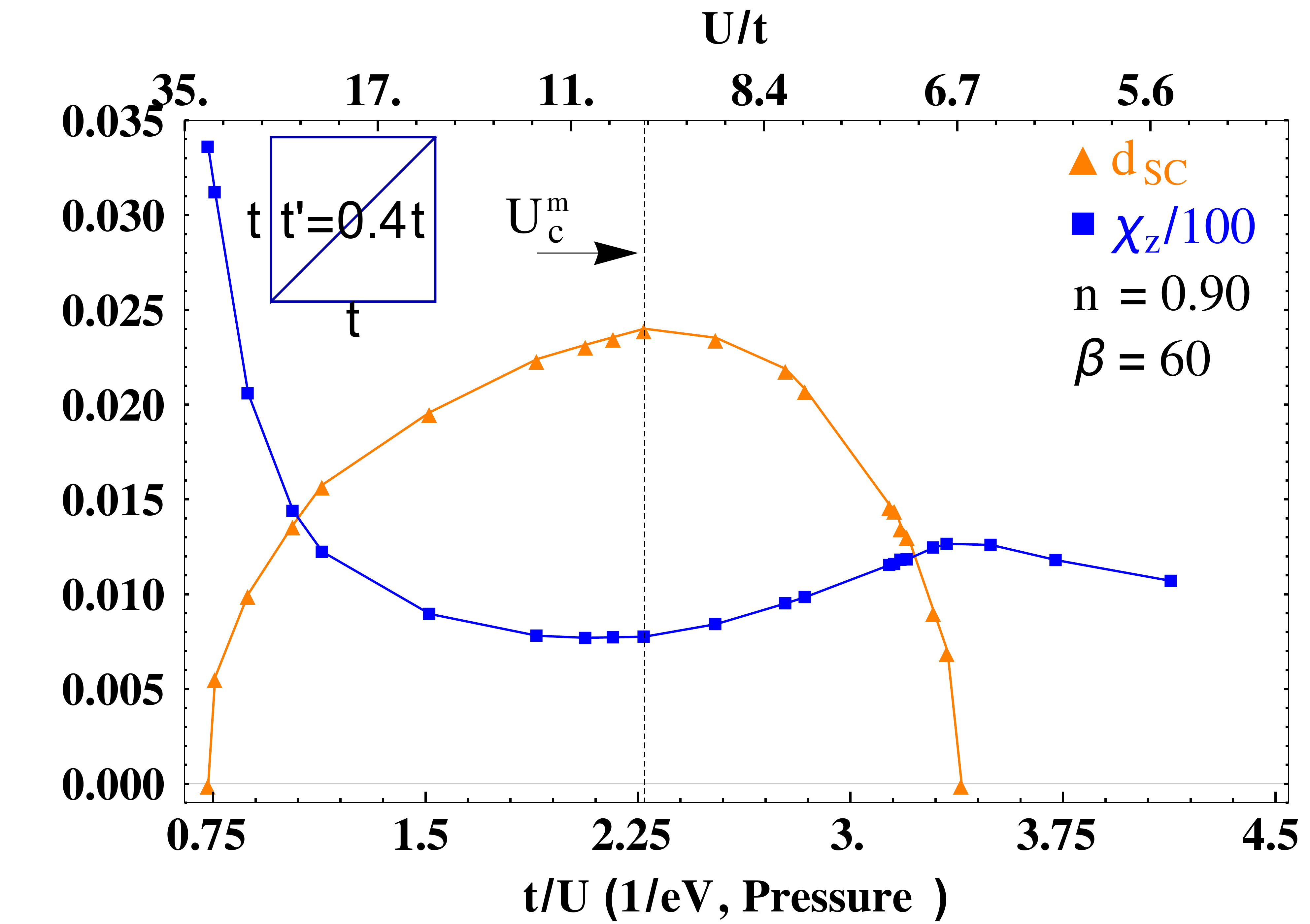}
		\label{fig:ChiSz-OrderParametertp0.4-n090}
	}
	
	\caption{
		(color online) Order parameter (\OP) and uniform susceptibility (\UCHI) for the superconducting phase of the Hubbard model on the anisotropic triangular lattice with $t'=0.4t$. Each filling has its specific color and plotmarker in the various panels of this figure and those of Fig.~\ref{fig:OrderParametertp0.8}.  Lines are guides to the eye.
		\newline
		\tg{(a)} \OP for the half-filled case. Two different temperatures are plotted.
		\newline
		\tg{(b)} \OP for various fillings. The  temperature is $T/t=1/60$. The dashed purple line is an extrapolation.
		\newline
		\tg{(c)}  \OP (green diamonds) and  \UCHI (blue squares) for the $1\%$ hole-doped case. \UCHI is divided by $100$ to fit on the same vertical axis. $U_{MMT}$ denotes the vicinity of interaction strengths where the pseudogap to metal first-order transition  discussed in Sec.~\ref{Sec:normal} is found. $U_{c}^{\, m}$ stands for the critical interaction strength associated with \TCM, the maximum $T_c$.
		\newline
		\tg{(d)} \OP (orange triangles) and \UCHI (blue squares) for the $10\%$ hole-doped case. \UCHI is divided by $100$ to fit on the same vertical axis. $U_{c}^{\, m}$ has the same significance as in (c). The orange curve for \OP is the same as in (b).
	}
	\label{fig:ChiSz-OrderParametertp0.4}
\end{figure*}

\paragraph*{The SC State} 
Fig. \ref{fig:PlaquetteDopageDiagramtp04}  illustrates the dramatic effects of doping on the SC phase. 
For  $1\%$ hole-doping (Fig.~\ref{fig:PlaquetteTp04-n099}), the SC dome is extended by a factor of six on the pressure axis at $T/t=1/60$ compared with the half-filled case shown in  Fig.~\ref{fig:PlaquetteTp04-n100}, while \TCM (the maximum $T_c$) is enhanced by approximately $25\%$. Suppressing the Mott insulating phase by doping allows the SC state to extend its stability far beyond the critical interaction strength for the Mott transition at half-filling ($U_{MIT}$) . Comparison of Fig.~\ref{fig:PlaquetteTp04-n099} and Fig.~\ref{fig:PlaquetteTp04-n090}, reveals that increasing hole-doping moves \TCM  to lower pressures. At 10\% hole doping, \TCM decreases slightly (by about 1\%) compared to the half-filled case.  By contrast, Fig.~\ref{fig:PlaquetteTp04-n110} shows that electron-doping (10\%) displaces  \TCM to significantly lower temperature compared to half-filling (about 15\%).
The value of \TCM as a function of doping appears in Table \ref{table:SC}.
As suggested by this Table and Figs.~\ref{fig:PlaquetteTp04-n099}, \ref{fig:PlaquetteTp04-n090}, the maximum doping for \TCM is found for intermediate hole-doping at a value of \UU that is doping dependent. This is illustrated in Figs.~\ref{fig:TcMaximum} and~\ref{fig:3D}. Too much hole-doping ($10\%$) or electron-doping ($10\%$) effectively reduces \TCM. Nevertheless, every doping that we studied exhibits an enhancement of the range where the SC state appears on the pressure axis when compared with half-filling. 

\begin{table}[!htpb]
	\begin{tabular}{|c|c|c|}
		\hline
		\tg{n} & ($U_{c}^{\, m}/t$; $\beta_{c}^{\, m}) \, , \, t'=0.4t$ & ($U_{c}^{\, m}/t$; $\beta_{c}^{\, m}) \, , \, t'=0.8t$
		\\ \hline
		1.10 & ($8.3 \pm 0.2$; $29.0 \pm 0.2$) & -----
		\\ \hline
		1.00 & ($6.15 \pm 0.02$; $25.0 \pm 0.5$) & ($7.78 \pm 0.02$; $35.0 \pm 0.5$)
		\\ \hline
		0.99 & ($6.6 \pm 0.2$ ; $20.5 \pm 0.5$) & -----
		\\ \hline
		0.98 & ($6.8 \pm 0.2$ ; $18.9 \pm 0.1$) & -----
		\\ \hline
		0.97 & ($7.2 \pm 0.3$ ; $19.2 \pm 0.5$) & -----
		\\ \hline
		0.96 & ($7.3 \pm 0.2$ ; $19.5 \pm 0.5$) & -----
		\\ \hline
		0.94 & ($8.0 \pm 0.5$ ; $20.8 \pm 0.3$) & -----
		\\ \hline
		0.92 & ($9.0 \pm 0.5$ ; $22.5 \pm 0.5$) & -----
		\\ \hline
		0.90 & ($10.5 \pm 0.5$; $25.2 \pm 0.2$) & ($14.5 \pm 0.5$ ; $37.0 \pm 0.5$)
		\\ \hline
	\end{tabular}
	\caption{Estimated CDMFT values of the maximum superconducting transition temperature and corresponding interaction strength ($U_{c}^{\, m}/t$; $\beta_{c}^{\, m}$) ($T_{c}^{\, m}=1/\beta_{c}^{\, m}$) for various fillings and two values of  frustration $t'/t$.  
	}
\label{table:SC}
\end{table}


\paragraph*{} 
The general aspect of the dome near the triangular right part of  Figs.~\ref{fig:PlaquetteTp04-n100} and \ref{fig:PlaquetteTp04-n099} shows that this section of the SC state for the $1\%$ hole-doped case is continuously connected to the SC phase at half-filling. For the range of \UU that is insulating at half-filling, the SC transition temperature as a function of doping at fixed $U$ seems however to vanish extremely steeply between $1\%$ and $0\%$ doping, by analogy with  cuprates.~\cite{Frattini:2015,Sordi_Strong_2012,Semon_Ergodicity_2014} Larger cluster calculations for the cuprates~\cite{gull_superconductivity_2013} suggest the existence of a smooth maximum. The overall qualitative shape of the SC phase diagram as a function of $T-U-n$ shown in Fig.~\ref{fig:3D} combines the results of Fig.~\ref{fig:PlaquetteDopageDiagramtp04}. The line of \TCM from Fig.~\ref{fig:TcMaximum} also appears on this plot. The constant doping scans of this plot are complementary to the constant \UU scans of Refs.~\onlinecite{Sordi_Metal_Transitions_2011,Sordi_Finite_Doping_2010}. 

The order parameter and  the uniform susceptibility in the SC state (discussed below) illustrate in more detail the evolution from half-filling to finite doping.

\paragraph*{Order parameter and magnetic susceptibility in the SC state} 
Fig.~\ref{fig:ChiSz-OrderParametertp0.4} displays the d-wave SC order parameter (\OP) calculated using Eq.~\ref{eq:ParametreSupra} for $t'=0.4t$. As seen in Fig.~\ref{fig:OrderParametertp0.4-Half-filling} for half-filling, \OP is largest near the first-order transition to the insulating phase. This confirms previous $T=0 $ results ~\cite{Kyung_Mott_2006}. Also, as one would expect, as the temperature is raised, the magnitude of \OP and the range of pressure where it is non-zero decrease. Furthermore, \OP obtained at $T=0$ in Ref.~\cite{Kyung_Mott_2006} and at low $T$ in Fig.~\ref{fig:OrderParametertp0.4-Half-filling} follows qualitatively the same pressure dependence as $T_c$ obtained here. 

Fig. \ref{fig:OrderParametertp0.4-All} indicates that the pressure dependence of \OP at low temperatures also follows qualitatively that of $T_c$ for all other fillings studied, except for the $1\%$ hole-doped case, where an anomaly is present near the pseudogap to metal transition (Fig. \ref{fig:ChiSz-OrderParametertp0.4-n099}). This is analog with the unfrustrated square lattice where the doping dependence of the low $T$ value of \OP does not follow that of $T_c$ in the under-doped regime \cite{Sordi_Strong_2012,Sordi_c_axis_2013}.  

The pseudogap to metal transition leaves some traces in the SC state at $1\%$ doping through signatures in certain observables, such as the uniform susceptibility, $\chi_{z}(q=0,\omega=0) = $ \UCHI Eq.~\ref{eq:Susceptibility}. Indeed, as shown in Fig.~\ref{fig:ChiSz-OrderParametertp0.4-n099}, near the value $U=U_{MMT}$ where the transition between pseudogap and metal phases occurs in the normal state, the uniform susceptibility in the SC phase (blue squares) shows a large variation, suggesting a crossover in the fundamental metallic properties of the system, even in the SC state: over a small range of \UU (0.3), \UCHI varies by a factor of 4.3 when $U_{MMT}$ is crossed. Upon decreasing pressure, \OP (green diamonds) also starts to weaken. This last characteristic is reminiscent of the half-filled case, where \OP dies out at the first-order transition to the insulating phase
(Fig.~\ref{fig:OrderParametertp0.4-Half-filling}). Note by comparing Fig.~\ref{fig:PlaquetteTp04-n099} and Fig.~\ref{fig:ChiSz-OrderParametertp0.4-n099} that $T_c$ does not change as drastically as $\chi_{z}(q=0,\omega=0)$ or \OP upon crossing $U=U_{MMT}$ . The value $U_{c}^{\, m}$, corresponding to the maximum $T_c$, differs slightly from the value where \OP is maximum.

For the $10\%$ hole-doped case (and $10\%$ electron-doped case, not shown),  \OP is anti-correlated to \UCHI. This is illustrated in Fig.~\ref{fig:ChiSz-OrderParametertp0.4-n090}. This behavior is expected since it becomes more difficult to align the spin of the electrons along a magnetic field when more singlet pairs are formed. By contrast with the $1\%$ doping case, for $10\%$, the first-order pseudogap to metal transition is not found at accessible $T$ and \UU, so \UCHI shows no peculiar behavior. 

\paragraph*{} 
The $T=0$ extrapolations of our results are consistent with the variational Monte Carlo results of Watanabe \textit{et al.}~\cite{Watanabe_SC_2014} for the same model. In the normal state they find a rapid crossover between two different metallic states  at finite-doping. They also find that the SC phase for $t'=0.4t$ is stable between \UU = 5 and \UU = 30 for a hole-doping of 8.3\%. Our numbers are \UU =  6.70 and \UU = 30 for 10\% hole-doping at $T/t = 1/60$.


\subsection{$t' = 0.8t$}
\label{subsec:Tp0.8}

\begin{figure}[]
	\centering
	\subfigure[]
	{
		\includegraphics[width=\linewidth]{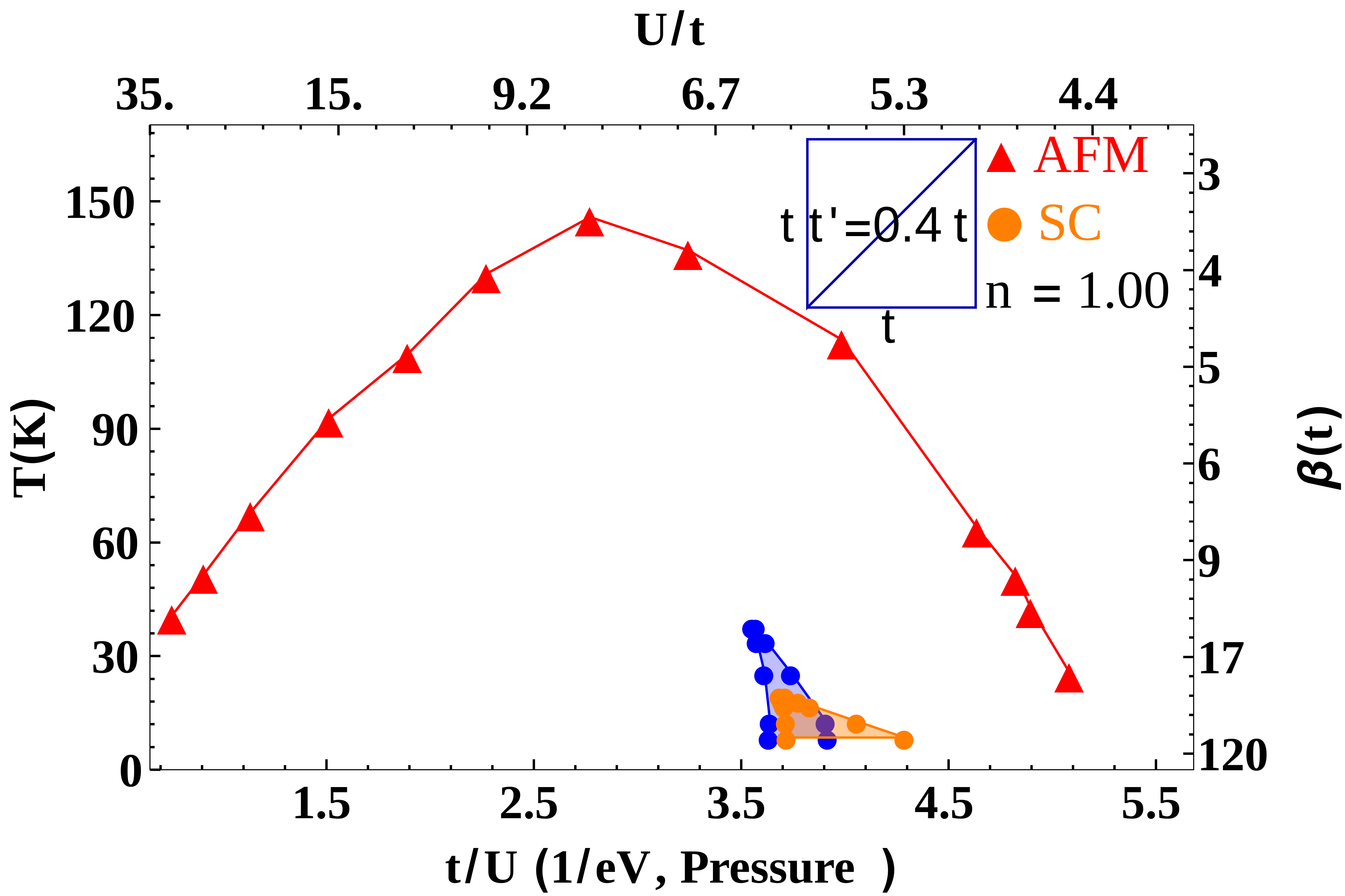}
		\label{fig:PlaquetteHalfFillingTp0.4_AFM_Total}
	}
	\newline
	\subfigure[]
	{
		\includegraphics[width=\linewidth]{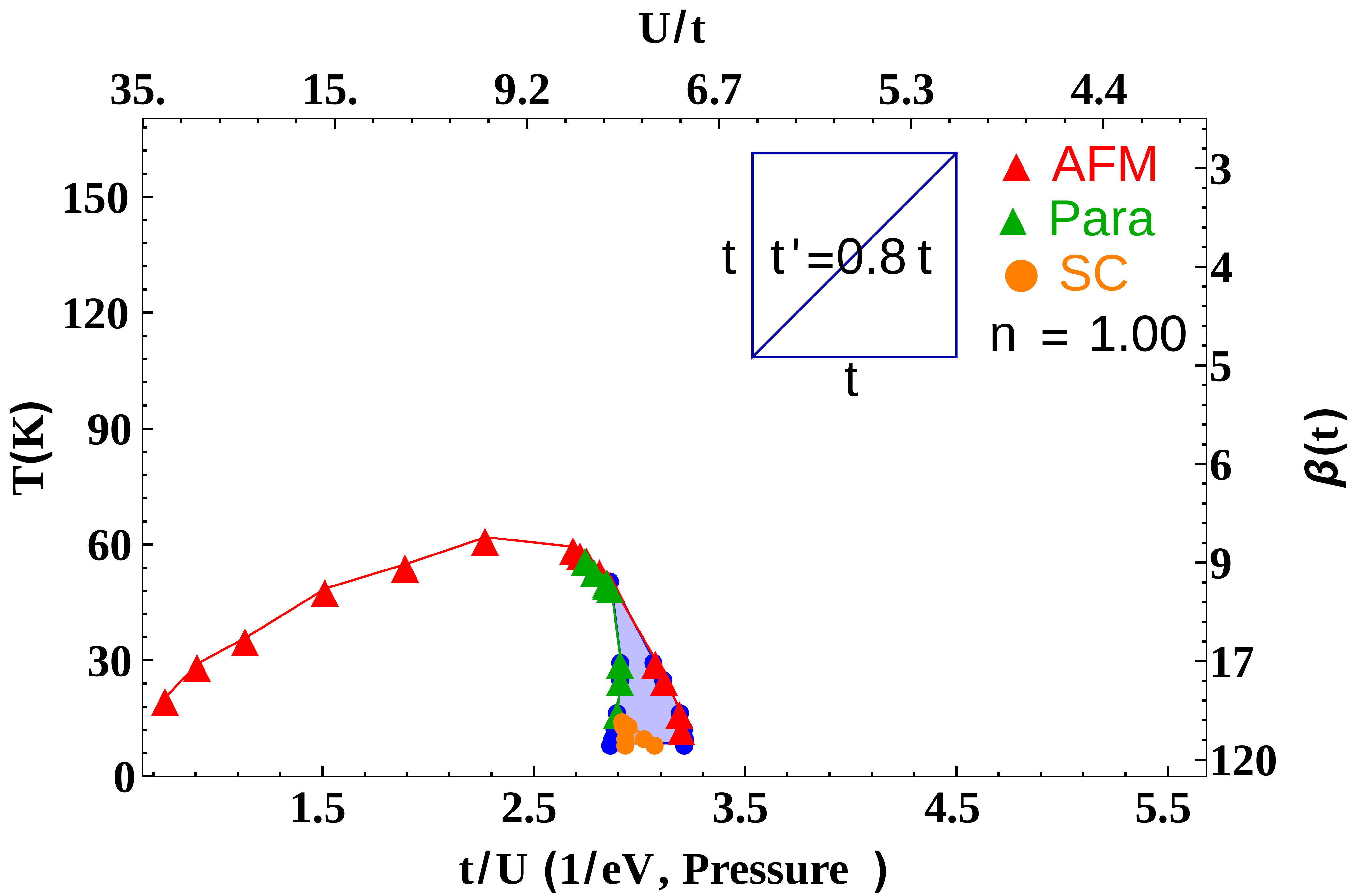}
		\label{fig:PlaquetteHalfFillingTp0.8_AFM_Total}
	}
	\caption{ 
		Phase diagrams for the Hubbard model on the anisotropic triangular lattice. $\beta$ is the inverse temperature in $1/t$ units. The value  $t=0.044$ eV is used to convert to physical units.~\cite{Liebsch_Mott_Triangle_2009} Lines are guides to the eye. \\
		\tg{(a)} Phase diagram at half-filling for an anisotropy parameter $t'=0.4t$. This figure is the same as Fig. \ref{fig:PlaquetteTp04-n100} except that it shows the complete antiferromagnetic phase.
		\newline
		\tg{(b)} Phase diagram at half-filling for an anisotropy parameter  $t'=0.8t$. This figure is the same as Fig. \ref{fig:PlaquetteHalfFillingTp0.8} except that it shows the complete magnetic phase.
	}
	\label{fig:Plaquette_AFM_Total}
\end{figure}

\begin{figure}[!htbp]
\centering
\subfigure[]
{
\includegraphics[width=\linewidth]{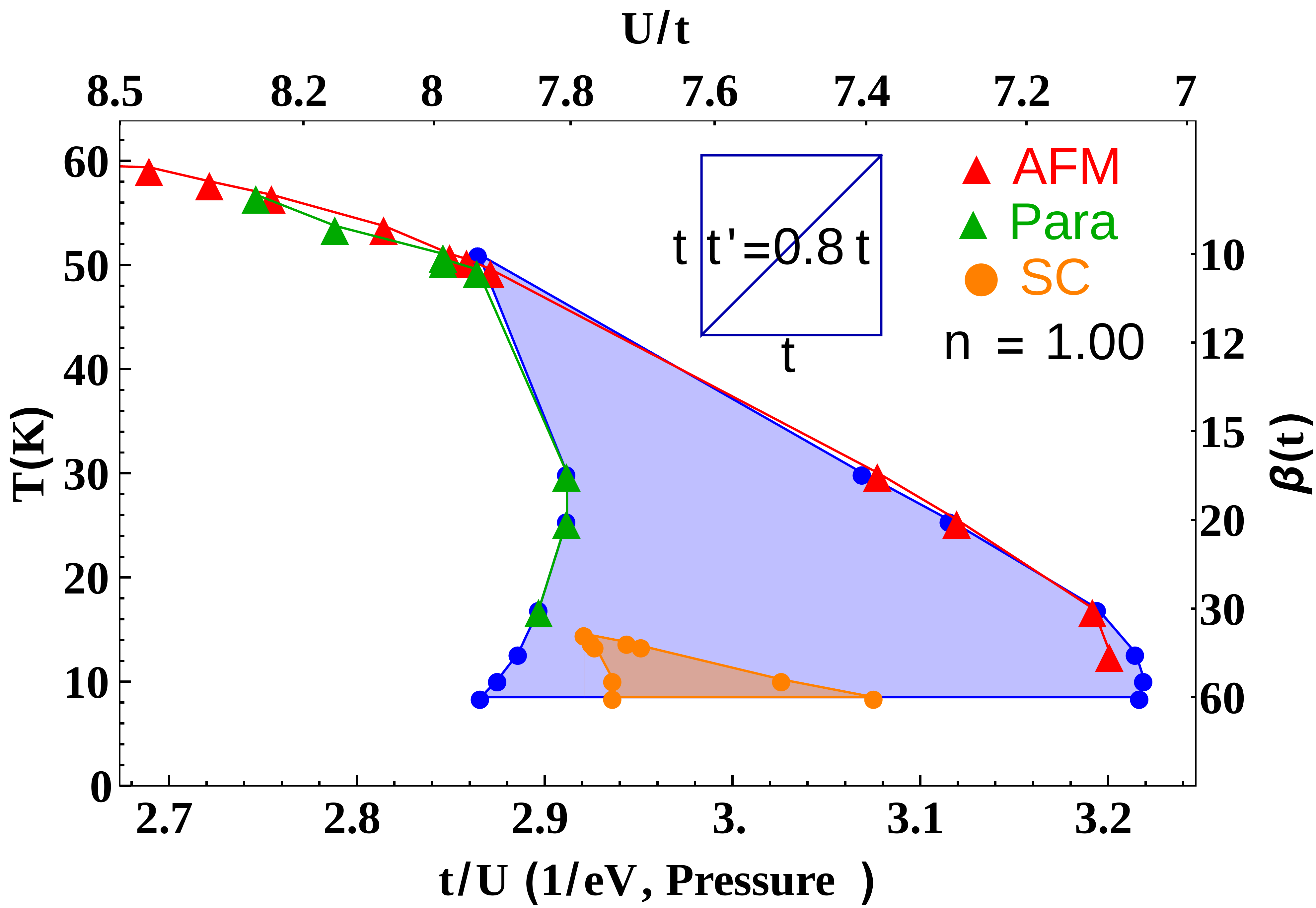}
\label{fig:PlaquetteHalfFillingTp0.8}
}
\newline
\subfigure[]
{
\includegraphics[width=\linewidth]{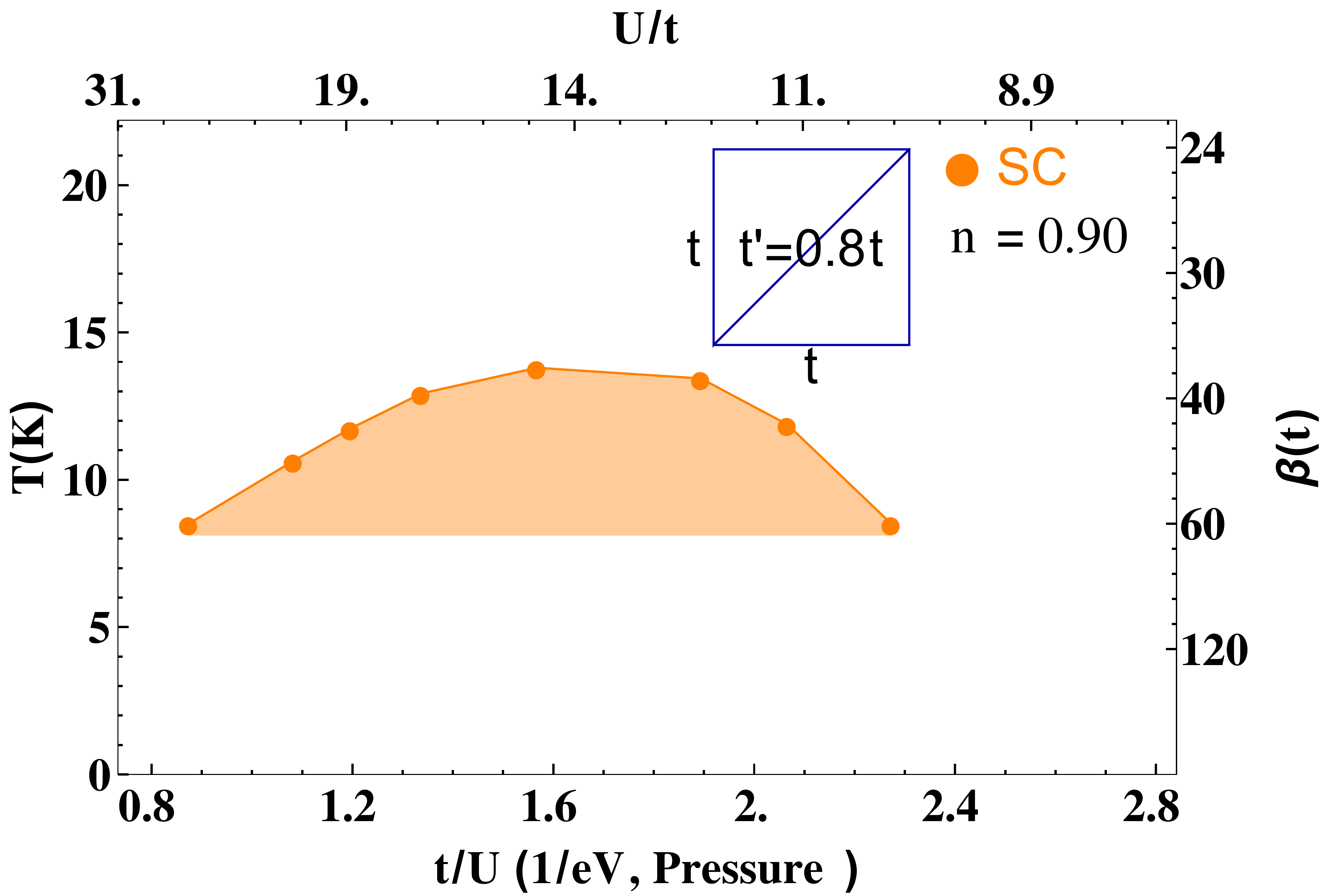}
\label{fig:PlaquetteTp08-Dopage}
}
\caption{ 
(color online) Phase diagrams for the Hubbard model on the anisotropic triangular lattice with anisotropy parameter $t'=0.8t$ and two fillings. The antiferromagnetic state was studied at half-filling and also for $n=0.9$.  In the latter case,  no AFM was found for $T$ down to $1/60$ and $8<U/t<30$.  The value  $t=0.044$ eV is used to convert to physical units.~\cite{Liebsch_Mott_Triangle_2009} Lines are guides to the eye. \\
\tg{(a)} Phase diagram for the half-filled case. Metallic and insulating phases coexist in the blue region. A first-order jump in the normal-state double occupancy is found at the blue points. The value of the critical point for the Mott transition is (\UU $=7.93$, $\beta=10.00$). With increasing pressure, we find at the red triangles that the N\'{e}el AFM order parameter disappears (to the right of the red line) through a first-order jump for $\beta > 10$ and through a second-order transition for $\beta < 10$.  When pressure is decreased, the paramagnetic metal is unstable to the AFM insulator along the green line through a first-order jump ($\beta > 10$) or through a second order transition ($\beta < 10$). The orange region is where superconductivity manifests itself. The SC state gives way to the insulating phase along a first-order jump upon decreasing pressure, and to the metallic phase along a second-order transition line upon increasing pressure. 
\newline
	\tg{(b)} Phase diagram for the $10\%$ hole-doped case. The transition from the SC state to the normal state is second-order.
}
\label{fig:PlaquetteDopageDiagramtp08}
\end{figure}

\paragraph*{} 

We move to the case $t^\prime=0.8t$, contrasting the results with the less frustrated case $t^\prime=0.4t$ just considered. We expect that for larger frustration, both SC and AFM will be negatively affected, leaving more room to effects related to the Mott transition. However, as we shall see, the effects of frustration are much stronger on the AFM than on the SC and normal states.

For compounds with $t'=0.4t$, Fig.~\ref{fig:PlaquetteHalfFillingTp0.4_AFM_Total} shows that $T=0$ AFM order would mask the Mott transition, and leave either a small region of SC at large pressure and very low temperature, or coexisting AFM and SC phases. 

Figs.~\ref{fig:PlaquetteHalfFillingTp0.8} and \ref{fig:PlaquetteHalfFillingTp0.8_AFM_Total} at half-filling demonstrate that larger frustration, $t'=0.8t$, is more detrimental to AFM order than to SC order. The maximum N\'eel transition temperature for the AFM phase occurs at ($T/t \sim 1/3.5$, \UU$ \sim 8.2$) for intermediate frustration, $t'= 0.4t$ Fig. \ref{fig:PlaquetteHalfFillingTp0.4_AFM_Total}, whereas for $t'= 0.8t$ in Fig. \ref{fig:PlaquetteHalfFillingTp0.8_AFM_Total} one finds ($T/t \sim 1/8.25$, $U/t \sim 10$). Thus, the maximum AFM transition temperature is decreased by a factor of about 2.4. By contrast, the maximum $T_c$ decreases by about only 25 \%. This is the expected effect of frustration and it agrees qualitatively with FLEX calculations~\cite{Kino_Phase_Diagram_1998} . However, FLEX is not valid across a Mott transition. Concerning the nature of the AFM phase, $T=0$ studies with CDMFT~\cite{Kyung_Mott_2006} on $2\times 2$ cluster, find that between $t'/t=0.7$ and $t'/t=0.8$, superconductivity at the first-order transition changes from coexisting with a commensurate AFM phase to coexisting with a phase that is not magnetically ordered. Variational Monte Carlo studies find that commensurate AFM is stabilized for~\cite{Tocchio:2013}  $t'/t<0.75$ or even $t'/t<0.9$ ~\cite{Watanabe:2008}, consistent with our results. The latter early study however finds that magnetic states are always more stable than superconductivity.       

\paragraph*{} 
Long-wavelength AFM fluctuations are also detrimental to long-range order since the Mermin-Wagner-Hohenberg theorem requires that the N\'{e}el transition temperature ($T_{N}$) vanish in the absence of coupling to the third dimension. The $T_{N}$ lines that we find here are only indicators for the onset of the renormalized classical regime where low-frequency long-wavelength AFM fluctuations become important.  Furthermore, as discussed in \ref{subsec:Tp0.4}, hole-doping also suppresses $T_{N}$. 

The above considerations suggest that for $t'=0.8t$ our results for the normal state pseudogap to metal transition and for the SC phase are observable at finite temperature in real materials. The crossover discussed in  Secs.~\ref{subsec:Tp0.4} and \ref{Sec:normal} also occurs for $t'= 0.8t$  and is thus a fundamental feature of systems near half-filling. For $1\%$ hole-doping, we found the first-order transition at \UU$ \sim 8$. However, the pseudogap to metal transition and especially the SC phase were particularly difficult to study extensively due to a worse sign problem, hence they are not displayed. 

Results for the $10\%$ doped case are shown in Fig.~\ref{fig:PlaquetteTp08-Dopage}. The SC phase is present for a much broader range of pressure than in the half-filled case (Fig. \ref{fig:PlaquetteHalfFillingTp0.8}). Furthermore, the SC dome is shifted to lower pressure. These effects of hole-doping for $t'= 0.8t$ are very similar to those for $t'=0.4t$. Again, the effect of frustration on the superconducting $T_c$ is much smaller than on $T_N$. Indeed, $T_N$ changes from a finite value  at $t^\prime/t=0.4$ (Fig.~\ref{fig:PlaquetteDopageDiagramtp04}) to zero at $t^\prime/t=0.8$, while the maximum $T_c$ decreases only by about 30\%. 

\begin{figure}[!htbp]
\centering
\subfigure[]{
\includegraphics[width=\linewidth]{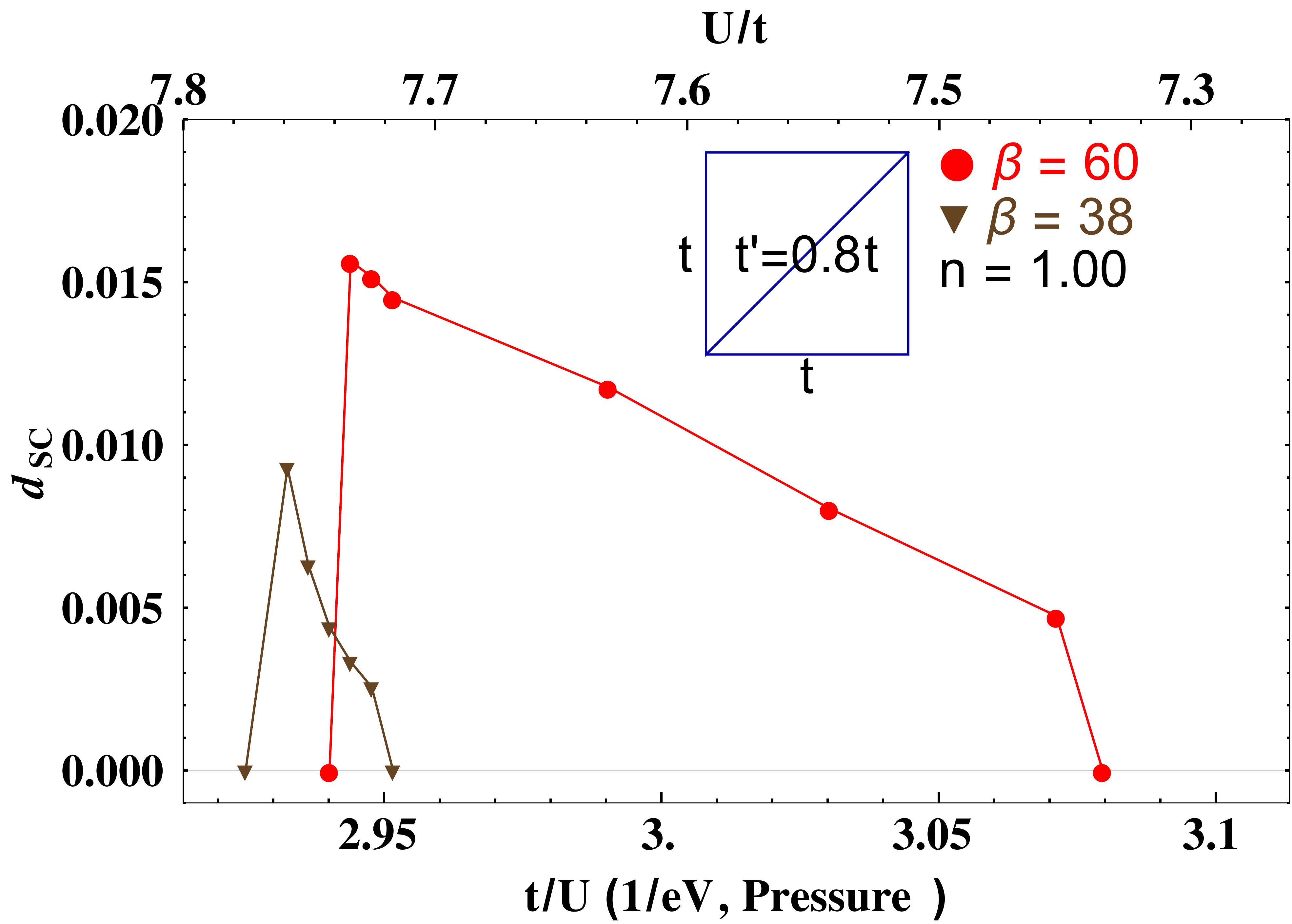}
\label{fig:OrderParametertp0.8-a}
}
\newline
\subfigure[]{
\includegraphics[width=\linewidth]{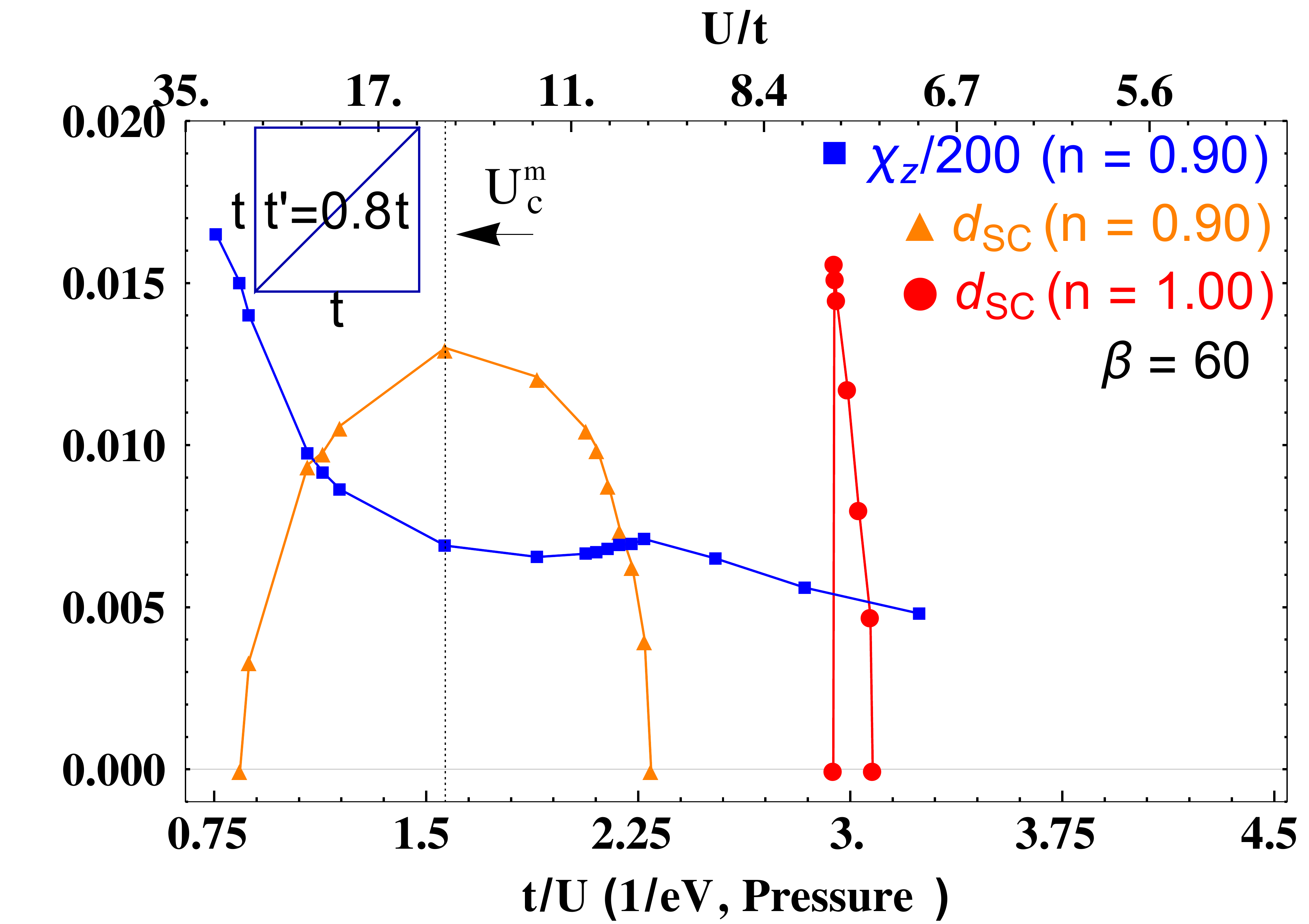}
\label{fig:OrderParametertp0.8-b}
}
\caption{
(color online) The order parameter (\OP) and uniform susceptibility (\UCHI) in the superconducting phase of the Hubbard model on the anisotropic triangular lattice with frustration value $t'=0.8t$. Each doping has the same specific color and plotmarker as in Fig.~\ref{fig:ChiSz-OrderParametertp0.4}. Lines are guide to the eye.
\newline
\tg{(a)} \OP for the  half-filled case. Two different temperatures are plotted.
\newline
	\tg{(b)} \OP  for various fillings and \UCHI for the 10\% hole-doped case. The temperature is $T/t=1/60$. $U_{c}^{\, m}$ stands for the superconducting critical interaction strength associated with \TCM. The spin susceptibility $\chi_z$ calculated with Eq.~\ref{eq:Susceptibility} in dimensionless units is divided by 200 to fit on the same vertical axis. 
}
\label{fig:OrderParametertp0.8}
\end{figure}

\paragraph*{SC order parameter} 
Fig.~\ref{fig:OrderParametertp0.8} displays \OP calculated with Eq. \ref{eq:ParametreSupra} for half-filling and for 10\% doping. The qualitative observations made for $t'=0.4t$ for 10\% doping still hold here at larger frustration, namely the pressure dependence of \OP  at low temperatures follows qualitatively that of $T_c$ and doping increases drastically the range of the SC dome on the pressure axis. Here too, the uniform susceptibility is anti-correlated to \OP. 

\subsection*{Additional comparisons with earlier results} 
The range of \UU where the SC phase appears for 10\% hole-doping and $t'=0.8t$, namely $10 <  U/t < 26$ for $T/t = 1/60$, is similar to that found~\cite{Watanabe_SC_2014} with the variational Quantum Monte Carlo method  for  8.3\% hole-doping at $T=0$,  namely $10 < U/t < 25$. 

\paragraph*{} 
At half-filling, the shape of the coexistence region in our normal state phase diagram  differs from that found on a three site cluster with CDMFT with exact diagonalization solver by Liebsch \textit{et al}.~\cite{Liebsch_Mott_Triangle_2009}   Also, on the doped three-site cluster the pseudogap is not observed in the isotropic limit~\cite{KyungTriangle:2007}. These differences are not surprising since the ground state entropy of clusters with an odd or even number of sites is very different. The four site cluster results of Ohashi and \textit{al.} \cite{Ohashi_Finite_2008} for the transition line have a slope of the same sign as us in the $T-U$ plane but the actual values of $U$ and especially $T$ differ, a difference that may come from the systematic imaginary-time discretization errors of the Hirsch-Fye algorithm.

 \paragraph*{}
Our results extrapolated to $T=0$  are consistent with earlier  CDMFT results obtained with an exact-diagonalization impurity solver.~\cite{Kyung_Mott_2006} Here we also find with decreasing pressure that the system changes from a paramagnetic metal, to a superconductor to an AFM insulator.  In Ref.~\onlinecite{Kyung_Mott_2006}, the AFM phase boundary at sufficiently large frustration coincides with the Mott transition. This is consistent with the extrapolation to $T=0$ of the coincidence in Fig.~\ref{fig:PlaquetteHalfFillingTp0.8} between two transitions found with increasing pressure, namely the first-order jump between the AFM state and the paramagnetic metal and (when AFM long-range order is forbidden) the first-order jump between the Mott insulator and the paramagnetic metal. At smaller frustration, this does not occur: Upon decreasing pressure (or increasing $U$), the AFM transition occurs before the Mott transition, as can be seen for $t'/t=0.4$ in Fig.~\ref{fig:Plaquette_AFM_Total}. Also, as seen in Fig.~\ref{fig:Plaquette_AFM_Total}, the critical value of $U/t$ for the Mott transition at half-filling increases with frustration $t^\prime/t$, in agreement with Refs.~\cite{Parcollet:2004,Kyung_Mott_2006}.  This is reflected in the qualitative shape of the magenta region in Fig.~\ref{fig:GeneralizedPhaseDiagram}. The Mott transition on the anisotropic triangular lattice has been studied with many other methods, for example path integral renormalization group~\cite{MoritaImadaMott:2002}. The critical value of $U/t$ found with the latter method is smaller than that in CDMFT. Note finally that it has been found earlier in different contexts that depending on frustration $t'/t$, AFM transitions can be of first or second order \cite{Zeng_Collinear_AFM, Kondo_AFM_First_Order, Hofstetter_AFM_First_Order}.

\paragraph*{} 
The $T=0$ magnetic phases found for various values of $t'/t$ and $U/t$ with the variational cluster approximation~\cite{Sahebsara_Hubbard_2008, Laubach_Competing_Magnetism,Yamada_Magnetic_Triangle_2014} exact diagonalization of an effective model~\cite{yang_effective_2010}, variational Monte Carlo,~\cite{Tocchio:2013,Watanabe:2008}  and dual-fermions~\cite{Laubach_Competing_Magnetism} are similar to those found with CDMFT in Ref.~\onlinecite{Kyung_Mott_2006}, except that here and in Ref.~\onlinecite{Kyung_Mott_2006} the possibility of spiral order has not been investigated. Nevertheless, according to Refs.~\cite{Tocchio:2013,Thomale:2011,Watanabe:2008}, commensurate AFM of the type we find can be stable up to $t'/t\sim 0.8$. In ~\cite{Watanabe:2008} superconductivity is always less stable than AFM but this is not so in weak correlation approaches such as FLEX~\cite{Kino_Phase_Diagram_1998} or functional renormalization group~\cite{MarstonRG:2001}. With the latter approach, superconductivity is also obtained for $t'/t=1$ but symmetry considerations in this case are different.~\cite{HonerkampTriangle:2003} At half-filling, it has been widely appreciated for a long time by many methods that superconductivity is stable for a wide range of values of $t'/t$: in variational~\cite{LiuVariationalTriangle:2005,Watanabe:2006} CDMFT~\cite{Kyung_Mott_2006} Gutzwiller~\cite{GanGutzwiller:2005} resonating valence bond~\cite{PowellRVB:2005} approaches. 

\section{Discussion}\label{sec:discussion}


\subsection{Broken-symmetry phases}\label{sec:discussion_broken_symmetry}
\paragraph*{} 
Although phase transitions in CDMFT are renormalized by local dynamical fluctuations, they essentially have a mean-field character. In particular, they do not satisfy the Mermin-Wagner-Hohenberg theorem that forbids continuous-symmetry breaking at finite temperature in two dimensions.~\cite{Mermin_Theorem} This is especially relevant for AFM order, which we overestimate. On the other hand, the superconducting critical temperature $T_c$ found in our phase diagrams physically represents the dynamical mean-field transition temperature $T_c^d$ below which Cooper pairing occurs locally in the cluster.~\cite{Sordi_Strong_2012} Long-wavelength quantum and thermal fluctuations in the amplitude and phase of the order parameter \OP should lead to an actual Kosterlitz-Thouless  transition temperature smaller  than  $T_c^d$ \cite{Ussishkin_SC_Fluctuations,Podolsky_SC_Nernst,Tesanovic_d-wave_2008} With increasing cluster sizes, the dynamical mean-field $T_c$ have been shown to converge to a finite value on the square lattice.~\cite{Maier_Cluster_Size_SC}  

\subsection{Strongly correlated superconductivity, superconducting dome and AFM quantum critical point}

The link between antiferromagnetic quantum critical point and unconventional superconductivity is well documented, especially in the field of heavy-fermion materials.~\cite{yang_quantum_2014} Numerical simulations with methods very close to those used here confirm this intimate connection for the Anderson lattice model of heavy-fermions:~\cite{wu_d-wave_2014}  Indeed, one finds a superconducting dome that systematically surrounds the antiferromagnetic quantum critical point. The same type of connection to a quantum-critical point has been proposed for cuprates.~\cite{ramshaw_quantum_2014,taillefer_scattering_2010}  We suggest that this connection between AFM quantum critical point and maximum $T_c$ is present when the interaction strength is not large enough to lead to a Mott transition. In that case pairing occurs through the exchange of long wavelength antiferromagnetic fluctuations.~\cite{Beal-Monod:1986,Scalapino:1986,Bickers_dwave:1989,MaierMacridinJarrellScalapino:2007} 

The top panel of Fig.~\ref{fig:Plaquette_AFM_Total} shows that for half-filled organics, where a Mott transition is clearly observed, superconductivity is near the Mott transition, not near the antiferromagnetic quantum critical point. And as we dope, Fig.~\ref{fig:PlaquetteTp04-n099} shows that the superconducting dome surrounds the pseudogap to metal transition that is the finite-doping remnant of the Mott transition. 

\begin{figure}
	\includegraphics[width=\linewidth]{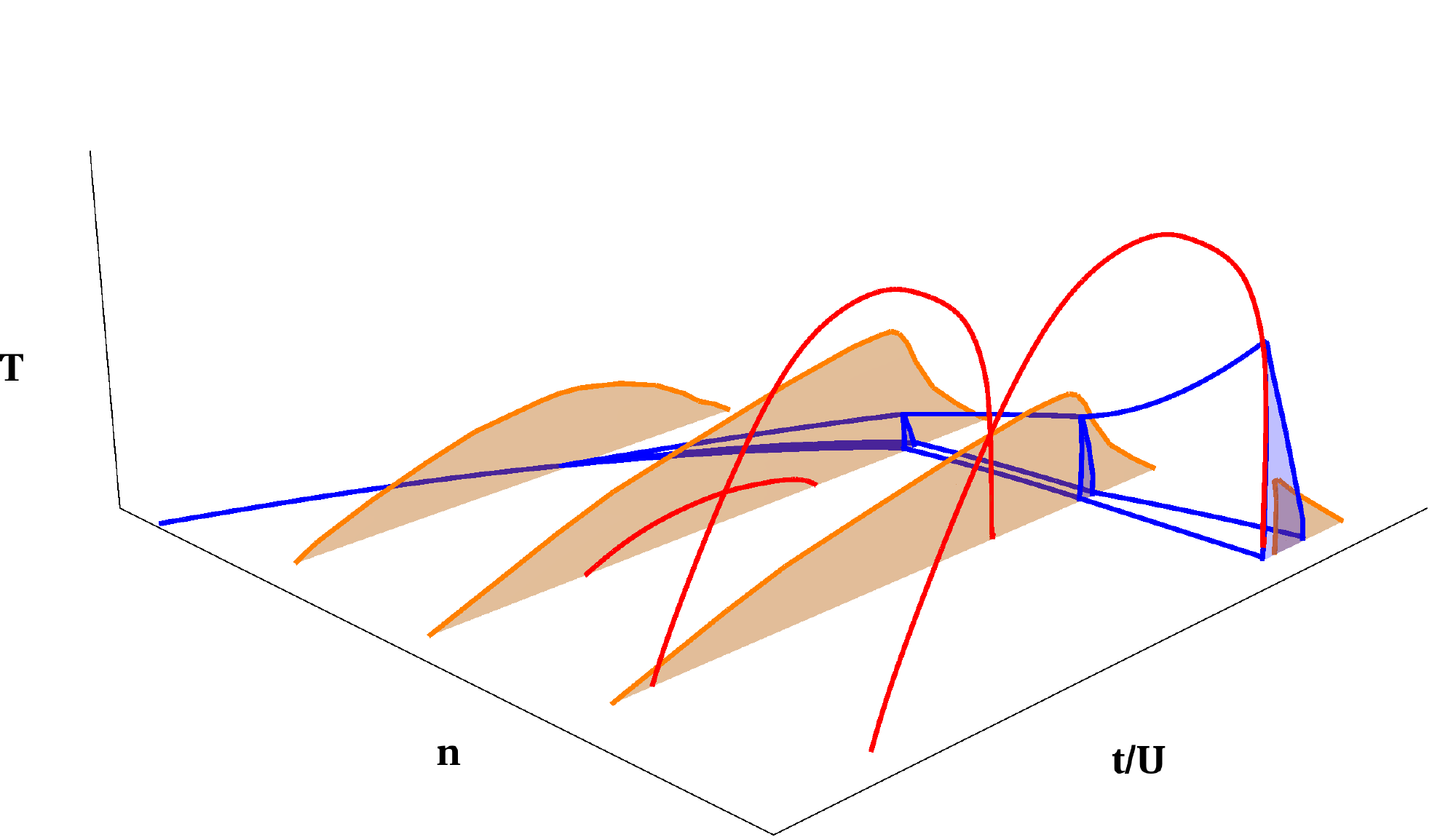}
	\caption{Schematic phase diagram for layered organics. Orange planes indicate the three-dimensional region where superconductivity is present. The antiferromagnetic three-dimensional dome is below the red lines. In the normal state, there is a surface of first-order transition whose coexistence region is delimited by the blue lines. This first-order transition separates a pseudogap phase with small double-occupancy from a correlated metal with larger double-occupancy, by analogy with Ref.~\onlinecite{Sordi_Finite_Doping_2010}. At larger $U$ and doping, the sign problem prevents us from observing directly the first-order transition or its continuation as a different phenomenon. Frustration can appreciably displace and eventually completely eliminate the antiferromagnetic region.~\cite{Kyung_Mott_2006,Sahebsara_antiferromagnetism_2006}}
	\label{fig:schematic}
\end{figure}

A schematic phase diagram  is displayed in Fig.~\ref{fig:schematic}.  In the normal state, there is a first-order phase transition whose coexistence region is represented in blue. The maximum of the superconducting transition temperature is controlled by the opening of the pseudogap, namely by the position of the first-order phase transition or its continuation, not by the antiferromagnetic quantum critical lines at the end of the antiferromagnetic three-dimensional dome delimited by the red region: even in the absence of the antiferromagnetic phase, superconductivity survives, as can be verified from Fig.~\ref{fig:PlaquetteDopageDiagramtp08}b. This is a characteristic of strongly-correlated superconductivity in doped Mott insulators. We stress however that at large doping we observe crossovers in the normal state, but we cannot calculate at low enough temperature to confirm if the first-order transition survives. It can in principle be replaced by a $T=0$ second order transition line or disappear at a critical point. 

The mechanism for superconductivity in the organics is thus clearly different from that associated with an antiferromagnetic quantum critical point. This type of strongly-correlated superconductivity is controlled by short-range AFM correlations, namely superexchange $J=4t^2/U$, as found early on in slave-boson calculations~\cite{Kotliar:1988} and more recently with CDMFT~\cite{Kancharla:2008}. This is reviewed in Ref.~\onlinecite{TremblayJulichPavarini:2013}.

Further studies will be needed to clarify the detailed cause of the $T_c$ dome. Consider the optimal $T_c$ in Figs.~\ref{fig:PlaquetteTp08-Dopage},	\ref{fig:PlaquetteTp04-n099}, \ref{fig:PlaquetteTp04-n090}, \ref{fig:PlaquetteTp04-n110}. The decrease of $T_c$ from optimal towards low pressure (on the left) scales like $J$ (i.e. like a straight line). On the other hand, the opening of a pseudogap could also explain this decrease because a pseudogap in the density of states at the Fermi level leaves fewer states to pair. 

\subsection{Contact with experiment}

\paragraph*{} 
Our  phase diagrams at half-filling (Figs.~\ref{fig:PlaquetteTp04-n100}, \ref{fig:PlaquetteHalfFillingTp0.8}), 
reveal interesting similarities with the experimental phase diagrams of the BEDT compounds but even more so with the \DMITS compounds.~\cite{Shimizu_Dmit, Yamaura_Dmit}  In the temperature-pressure plane, the shape of the region where the SC phase exists and the shape of the metal-insulator coexistence region  are in qualitative agreement with experiment. The maximum SC transition temperature \TCM as a function of pressure coincides with the phase transition to the insulator, also like in experiment. 

\paragraph*{} 
We find that as frustration is increased, both the maximum $T_N$ and \TCM decrease, the decrease in the tendency to magnetic order being however much more pronounced. These observations are consistent with experiments where compounds with higher frustration $t'/t$ have lower $\TC$ compared to less frustrated materials. For example, \KCN ($t'=0.8t$) and \KCl ($t'=0.4t$) display \TCM of respectively 3.9K and 13.1K. Also, as frustration is increased in real compounds, AFM order is greatly suppressed. For instance, \KCl has AFM order at low temperatures and pressures, while \KCN is a spin-liquid candidate.~\cite{Lefebvre_Phase_Diagram_2000,Shimizu_SpinLiquid_2003}

\paragraph*{} 
Fig. \ref{fig:PlaquetteDopageDiagramtp08} at $t'=0.8t$ accounts for the experimental results on the SC phase both at half-filling \cite{Shimizu_Pressure_Induced_2010} and at $10 \%$  doping.~\cite{Oike_Mottness_2014} Indeed, in half-filled \KCN a superconductor-insulator phase transition is observed upon decreasing pressure, as in Fig.~\ref{fig:PlaquetteHalfFillingTp0.8}. On the other hand $\kappa$-(ET)$_{4}$Hg$_{2.89}$Br$_{8}$, the $10\%$ hole-doped counterpart of \KCN, presents a dome shaped~\cite{Oike_Mottness_2014,taniguchi_anomalous_2007} $T_c$ similar to Fig. \ref{fig:PlaquetteTp08-Dopage} except that the dome is more asymmetrical in experiment. 

Fig. \ref{fig:PlaquetteDopageDiagramtp08} for $t'=0.8t$ also shows that the range of pressure where SC appears at $10\%$ hole-doping and $T/t$ = 1/60 is multiplied by about nine compared to the half-filled case. Rough extrapolation of the superconducting dome to $T=0$ gives a range of pressure that increases by a factor four to six in going from half-filling to 10\% doping. Experimentally the dome is  extended  by approximately six for the same  value of frustration. Indeed, in half-filled \KCN, superconductivity occurs over a range of 0.25~GPA according to Ref.~\cite{Kurosaki_kCN_Phase_diagram} while Ref.~\cite{Oike_Mottness_2014} finds superconductivity for a range of 1.5~GPa in the doped compound $\kappa$-(ET)$_{4}$Hg$_{2.89}$Br$_{8}$. 

\TCM for \KCN is $3.9K$ while it is about $7K$ for the doped counterpart, a factor of $1.8$. Our results show a very slight decrease of \TCM  at $10\%$ hole-doping. However, \TCM is  increased for intermediate doping (Table \ref{table:SC}, Fig.~\ref{fig:TcMaximum}) compared to  half-filling. While that increase for intermediate doping is proven for an anisotropy parameter $t'=0.4t$, the case $t' = 0.8t$ should be similar. 

Oike \textit{et al.}~\cite{Oike_Mottness_2014} and Taniguchi \textit{et al.}~\cite{taniguchi_anomalous_2007} also found in the Hall coefficient of $\kappa$-(ET)$_{4}$Hg$_{2.89}$Br$_{8}$ a rapid crossover around $0.5$ GPa.  This pressure corresponds closely to the maximum of the superconducting transition temperature~\cite{Oike_Mottness_2014} \TCM.  Although for $t'/t=0.8$ and $n=0.9$ we cannot reach low-enough temperature to detect the first-order pseudogap to metal transition, the $t'/t=0.4$ low-doping results of Fig.~\ref{fig:PlaquetteTp04-n099}  strongly suggest that \TCM is controlled by that transition. On the square lattice, one finds analogous results:~\cite{Sordi_Strong_2012} A low temperature first-order transition between a pseudogap metallic phase and a strongly-correlated metal ends at a critical point above which a line of crossovers appears.~\cite{Sordi_Widom,Sordi_c_axis_2013} This line of crossovers is a Widom line, a general phenomenon found in the supercritical region of first-order transitions.~\cite{XuStanleyWidom:2005}  Remarkably, \TCM is near the intersection of the superconducting dome and of the Widom line.~\cite{Sordi_Strong_2012,Taillefer:2015} 
This leads us to important predictions for experiment. 

\subsection{Predictions}

\paragraph*{} 
The first-order pseudogap to metal transition, observed theoretically on the square lattice,~\cite{Sordi_Finite_Doping_2010,Sordi_Metal_Transitions_2011,Sordi_Widom} has also been seen as a sharp crossover in larger cluster calculations in the Dynamical Cluster Approximation~\cite{werner_momentum_2009} and in variational Quantum Monte Carlo.~\cite{yokoyama_crossover_2013} Our work shows that this transition also occurs on the anisotropic triangular lattice. It had been observed before as a sharp crossover.~\cite{Watanabe_SC_2014} The experimental results on the doped  BEDT $\kappa$-(ET)$_{4}$Hg$_{2.89}$Br$_{8}$~\cite{Oike_Mottness_2014,taniguchi_anomalous_2007} can thus be interpreted as an observation of the crossovers associated to this pseudogap to metal transition. Based on our phase diagrams, we predict that in yet to be synthesized very low-doping organic materials with pressure-induced transitions, remnants of this transition could be detectable in the SC state. It would manifest itself via observables such as the uniform susceptibility, the SC order parameter or  by a strong crossover of many properties as a function of pressure near \TCM.

We also predict that in electron-doped compounds, $\TC$ is  decreased  but the range of SC on the pressure axis is still increased compared to the half-filled case.

Finally, our results also suggest that for frustration high enough that magnetically ordered phases are absent, the normal state underlying the SC state of the {\it doped} compounds should display a first-order transition between a pseudogap and a more metallic state at sufficiently low temperatures. This is our most important prediction. As usual the normal state can be revealed by applying a magnetic field. Experiments {\it at half-filling}~\cite{KanodaBfield:2004} on $\kappa$-(BEDT-TTF)$_2$Cu[N(CN)$_2$]Cl  suggest that the magnetic fields necessary to destroy the SC state are easily accessible. 

\subsection{Limitations and perspectives}

The broad agreement that we find with experiment and, in the low temperature limit, with variational Monte Carlo methods,~\cite{Watanabe_SC_2014} suggest that the important physics in the layered organics arises from strong on-site repulsion $U$ and nearest-neighbor superexchange $J$. Increasing the cluster size would allow a better representation of the long-wavelength fluctuations beyond mean-field theory. It would produce more quantitative phase boundaries, as discussed in Sec.~\ref{sec:discussion_broken_symmetry} above. But since the existence of the superconducting phase itself has already been established by finite-size studies on the square lattice,~\cite{Maier_Cluster_Size_SC} it is highly unlikely that larger cluster studies would change this. 

The continuous time Quantum Monte Carlo impurity solver in the hybridization expansion that we have used here (CT-HYB) is for now the only Monte Carlo approach that allows calculations in the range of large \UU  and frustration needed for the layered organics. Recall that $U/t=14$ at the maximum of the superconducting dome for $t'/t=0.8$ and that $U/t$ is as large as $30$ at the lowest temperature end of the dome. Expansions in power of $U/t$ that are  used as Quantum Monte Carlo impurity solvers for larger clusters~\cite{Gull_Continuous-time_2011} fail for such large values of \UU and frustration because of a sign problem and because of the large expansion orders that are needed. Even on the 4-site cluster that we use, the average sign is immeasurably small with Rubtsov's algorithm for $U/t=14.5$ and $T/t=1/40$, namely near the maximum of the superconducting dome for $t'/t=0.8$.~\cite{Gull_Continuous-time_2011} For high-temperature, ($T/t \ge 0.06$) the cluster-size dependence has been shown to be negligible.~\cite{SakaiSize:2002} While CT-HYB provides a method to access the large values of  \UU and $t'/t$ that we need, the computation time increases exponentially with system size, making larger clusters unfeasible with present resources. The same size limitation applies to exact-diagonalization solvers~\cite{LiebschTong:2009} that, in addition, rely on a finite bath, by contrast with CT-HYB where the bath is infinite. A method has recently been proposed~\cite{Haverkort:2014} to increase bath size in exact diagonalization solvers but extensions to higher temperatures have not been tested and implementations on clusters have not been done yet.

Although AFM and SC were considered on equal footing, we neglected the possibility of non-vanishing order parameters for both AFM and SC simultaneously. This might occur in some regions at finite doping where we found that both AFM and SC separately can develop long-range order. The question of simultaneous non-vanishing order parameters is an interesting question but it is a detail at this early stage of investigations. The fact that for $t'/t=0.8$ model of the doped organic $\kappa$-(ET)$_{4}$Hg$_{2.89}$Br$_{8}$ there is a superconducting dome far from the AFM phase is one of the crucial proofs that the maximum of the dome does not come from an AFM quantum critical point. 

So far, studies of the Hubbard model have shown that they are capable of  capturing essential features of materials such as cuprates, \BEDT or \DMITS, even if that model neglects some physical effects.  Additional calculations taking into account Coulomb interaction between nearest neighbors (with the extended Hubbard model \cite{Tocchio_Phase_2014,Next_SC_Senchal}), electron-phonon interactions \cite{Bourbonnais_Phonons}, or the third spatial dimension \cite{Igoshev_Correlation_2014,Yoshikawa_Dimensionality}  would allow one to capture increasing details of these fascinating compounds, but the overall agreement with experiment that has been found suggests that the local and near-neighbor superexchange aspects of the Hubbard model capture the essential physics. 

Further investigations of the normal state pseudogap and of properties of the strongly correlated superconducting phase are planned. 

\section{conclusion}\label{sec:conclusion}
Based on CDMFT calculations of normal, superconducting and antiferromagnetic phase diagrams for the Hubbard model on the anisotropic triangular lattice, 
we arrive at the following conclusions.  
These phase diagrams are very similar to experimental observations, both at half-filling and at finite doping. Upon doping, superconductivity is enhanced; in particular it occurs over a much broader range of pressures (\UU). Smaller anisotropy, or equivalently larger frustration ($t'/t\sim 1$), diminishes antiferromagnetic and superconducting transition temperatures but antiferromagnetism is much more affected. 

In the normal state, a first-order pseudogap to metal transition occurs at finite doping and low temperature. The transition is continuously connected to the Mott transition at half-filling, as on the square lattice case,~\cite{Sordi_Finite_Doping_2010} yet the pseudogap is different from the Mott gap. The decrease in the density of states at the Fermi level associated with the pseudogap and the decrease of superexchange $J$ when pressure decreases can both contribute to the decrease of $T_c$ with decreasing pressure. We claim that competing antiferromagnetism is not an explanation for the dome in the doped organics. 

We predict that for very lightly hole-doped compounds, the pseudogap to metal transition leaves some subtle traces in the superconducting state. Our most important prediction is that the normal state that will be revealed by destroying the superconducting state with a magnetic field in lightly-doped highly-frustrated compounds will show this first-order transition between two conducting phases, one with a pseudogap and the other one metallic. It is this transition that should control the crossovers at finite temperature as well as the location of the maximal superconducting transition temperature. Finally, we predict that electron-doping should lead to a reduced maximum $\TC$. 

For a model of heavy fermions solved with the same set of methods as those used here, the maximum of the superconducting dome can be correlated with the location of an antiferromagnetic quantum critical point for interaction strengths that are not large enough to lead to a Mott transition.~\cite{wu_d-wave_2014} In our case, the superconducting dome in the doped organics surrounds the finite-doping extension of the zero-doping  first-order Mott transition (whenever it is directly observable). It is definitely not attached to an antiferromagnetic quantum critical point, a conclusion that can also be verified experimentally.   This result should clearly impact understanding of strongly-correlated superconductivity in all doped Mott insulators; not only layered organic superconductors, but also high-temperature cuprate superconductors.  

\begin{acknowledgments}

We are grateful to G. Sordi for numerous discussions, to Wei Wu for some calculations and to M. Gingras, C. Bourbonnais, and R. Nourafkan for suggestions and detailed comments on the manuscript. This work was supported by the Natural Sciences and Engineering Research Council of Canada (NSERC) and the Tier I Canada Research Chair Program (A.-M.S.T.). Simulations were performed on computers provided by CFI, MELS, Calcul Qu\'ebec and Compute Canada.

\end{acknowledgments}



%

\end{document}